\documentclass[12pt, preprint]{aastex}

\slugcomment{Accepted for publication in {\it The Astrophysical Journal}}

\shorttitle{X-Ray Variability of BL Lac objects}
\shortauthors{B\"ottcher \& Chiang}

\begin{document}

\title{X-ray spectral variability signatures of flares in BL Lac objects}

\author{Markus B\"ottcher\footnote{Chandra Fellow} \footnote{Current address:
Department of Physics and Astronomy; Ohio University; Athens, OH 45701}}

\affil{Department of Physics and Astronomy, Rice University,
6100 Main Street, Houston, TX  77005-1892}
\email{mboett@spacsun.rice.edu}

\and

\author{James Chiang}

\affil{NASA Goddard Space Flight Center, Code 661, 
Greenbelt, MD 20771 \\
Joint Center for Astrophysics and Physics Department,
University of Maryland, Baltimore, MD 21250}
\email{jchiang@elcapitan.gsfc.nasa.gov}

\begin{abstract}
We are presenting a detailed parameter study of the time-dependent
electron injection and kinematics and the self-consistent radiation
transport in jets of intermediate and low-frequency peaked BL~Lac
objects. Using a time-dependent, combined synchrotron-self-Compton 
and external-Compton jet model, we study the influence of variations 
of several essential model parameters, such as the electron injection 
compactness, the relative contribution of synchrotron to external soft 
photons to the soft photon compactness, the electron-injection spectral 
index, and the details of the time profiles of the electron injection 
episodes giving rise to flaring activity. In the analysis of our results,
we focus on the expected X-ray spectral variability signatures in a 
region of parameter space particularly well suited to reproduce the 
broadband spectral energy distributions of intermediate and low-frequency 
peaked BL Lac objects. We demonstrate that SSC- and external-Compton 
dominated models for the $\gamma$-ray emission from blazars are 
producing significantly different signatures in the X-ray variability, 
in particular in the soft X-ray light curves and the spectral 
hysteresis at soft X-ray energies, which can be used as a 
powerful diagnostic to unveil the nature of the high-energy
emission from BL~Lac objects.
\end{abstract}

\keywords{galaxies: active --- galaxies: jets --- radiation mechanisms:
nonthermal}

\section{\label{intro}Introduction}

The class of objects referred to as blazars consists of the most extreme 
examples of active galactic nuclei (AGNs), namely $\gamma$-ray loud,
flat-spectrum radio quasars (FSRQs), and BL~Lac objects. They have been 
observed in all wavelength bands --- from radio through very-high energy 
(VHE) $\gamma$-ray frequencies. More than 65 blazars have been identified 
as sources of $> 100$~MeV emission detected by the EGRET telescope on 
board the {\it Compton Gamma-Ray Observatory} (CGRO) \citep{hartman99}, 
and at least 6 blazars have now been detected at VHE $\gamma$-rays 
($> 350$~GeV) by ground-based air \v Cerenkov telescopes (for a 
recent review see, e.g., \cite{buckley01}). Blazars exhibit variability 
at all wavelengths \citep{vm95,mukherjee97} on time scales --- in some 
cases --- down to less than an hour \citep{gaidos96}. 

The broadband continuum spectra of blazars are dominated by
non-thermal emission and consist of at least two clearly 
distinct, broad spectral components. A sequence of sub-classes
of blazars can be defined through increasing peak frequencies 
and a decreasing dominance of the $\gamma$-ray output in terms
of $\nu F_{\nu}$ peak flux along a sequence from FSRQs via 
low-frequency BL~Lac objects (LBLs) to high-frequency peaked 
BL~Lac objects (HBLs), which is also correlated with a decreasing
inferred bolometric luminosity of the sources \citep{fossati98}.
For recent reviews of the observational properties of blazars
see, e.g., \cite{sambruna00}, \cite{pu01}, or \cite{boettcher02}.

Although all extragalactic sources detected by ground-based air
\v Cerenkov telescope facilities to date are HBLs, the steadily 
improving flux sensitivities and decreasing energy thresholds of 
those instruments provide a growing potential to extend their blazar 
source list towards intermediate and even low-frequency peaked BL~Lac 
objects. The detection of such objects at energies $\sim 40$ -- 100~GeV 
might provide an opportunity to probe the intrinsic high-energy 
cutoff of their spectral energy distributions (SEDs) since at those 
energies, $\gamma\gamma$ absorption due to the intergalactic infrared 
background is expected to be negligible at redshifts of $z \lesssim 0.2$ 
\citep{djs02}. Such detections should significantly further our
understanding of the relevant radiation mechanisms responsible for
the high-energy emission of blazars and the underlying particle
acceleration mechanisms.

The low-energy component of blazar SEDs is well understood
as synchrotron emission from ultrarelativistic electrons in
a relativistic jet directed at a small angle with respect to
the line of sight. In the framework of leptonic models (for
a review of the alternative class of hadronic jet models, 
see, e.g., \cite{rachen00}), high-energy emission will 
result from Compton scattering of lower-frequency photons off 
the relativistic electrons. Possible target photon fields for 
Compton scattering are the synchrotron photons produced within 
the jet (the SSC process; \cite{mg85,maraschi92,bm96}), or external 
photons (the EC process). Sources of external seed photons include 
the UV -- soft X-ray emission from the disk --- either entering the 
jet directly \citep{dsm92,ds93} or after reprocessing in the broad line 
region (BLR) or other circumnuclear material \citep{sikora94,bl95,dss97}
---, jet synchrotron radiation reflected at the BLR
\citep{gm96,bednarek98,bd98}, or the infrared emission from 
circumnuclear dust \citep{blaz00,arbeiter02}.

According to the now well-established AGN unification scheme
\citep{up95}, blazars can be unified with other classes of AGN, in
particular radio galaxies, through orientation effects. However, 
\cite{sambruna96} have pointed out that such orientation effects
can not explain the differences between different blazar sub-classes.
Instead, it has been suggested that the sequence of spectral properties 
of blazars from HBLs via LBLs to FSRQs can be interpreted in terms
of an increasing total power input into non-thermal electrons in
the jet, accompanied by an increasing contribution of external
photons to the seed photon field for Compton upscattering 
\citep{madejski98,ghisellini98}. It has been suggested that
this may be related to an evolutionary effect due to the gradual
depletion of the circumnuclear material being accreted onto the
central black hole \citep{dc00,cd02,bd02}. Detailed modeling of 
blazars in the different sub-classes (FSRQs, LBLs and HBLs) seems 
to confirm this conjecture (for a recent review, see, e.g.,
\cite{boettcher02}).

As mentioned earlier, blazars tend to exhibit rapid flux and
spectral variability. The variability is most dramatic and occurs 
on the shortest time scales at the high-energy ends of the two 
nonthermal spectral components of their broadband SEDs. Particularly 
interesting variability patterns could be observed at X-ray energies
for those blazars whose X-ray emission is dominated by synchrotron
emission. Observational studies of X-ray variability in blazars 
have so far focused on HBLs and, in particular, on the attempt 
to identify clear patterns of time lags between hard and soft 
X-rays. However, such studies have yielded rather inconclusive 
and often contradictory results (e.g., for Mrk~421: 
\cite{takahashi96,fossati00,takahashi00}; or PKS~2155-304:
\cite{chiapetti99,zheng99,kataoka00,edelson01}). Instead, 
the so-called ``spectral hysteresis'' of blazar X-ray spectral 
variability may prove to be a more promising diagnostic of the
physical nature of acceleration and cooling processes in blazar 
jets: When plotting the X-ray spectral hardness vs. the X-ray
flux (hardness-intensity diagrams = HIDs), some HBLs (e.g., Mrk~421
and PKS~2155-304) have been observed to trace out characteristic, 
clockwise loops (\cite{takahashi96,kataoka00}). In terms 
of pure SSC jet models, such spectral hysteresis can be 
understood as the synchrotron radiation signature of gradual 
injection and/or acceleration of ultrarelativistic electrons 
into the emitting region, and subsequent radiative cooling 
\citep{kirk98,gm98,kataoka00,kusunose00,lk00}. However, 
interestingly, such spectral hysteresis could not be confirmed 
in a recent series of {\em XMM-Newton} observations of Mrk~421
\citep{sembay02}. 

In LBLs, the soft X-ray emission is also sometimes dominated 
by the high-energy end of the synchrotron component 
\citep{tagliaferri00,ravasio02}, so similar spectral hysteresis 
phenomena should in principle be observable. However, those 
objects are generally much fainter at X-ray energies than their 
high-frequency peaked counterparts, making the extraction of 
time-dependent spectral information an observationally very
challenging task (see, e.g., \cite{bllac02}), which may 
require the new generation of X-ray telescopes such as 
{\em Chandra} or {\em XMM-Newton}. Extracting the physical 
information contained in the rich X-ray variability 
patterns exhibited by BL~Lac objects requires detailed 
theoretical modeling of the time-dependent particle 
acceleration and radiation transport processes in the 
jets of blazars. Previous analyses of these processes 
\citep{kirk98,gm98,cg99,kataoka00,kusunose00,lk00,krawczynski02}
have led to significant progress in our understanding of the 
particle acceleration and radiation mechanisms in HBLs, but
were restricted to pure SSC models, with parameter choices
specifically targeted towards HBLs. Consequently, those results may 
not be directly applicable to intermediate or low-frequency peaked
BL~Lac objects or even FSRQs. A notable exception is a recent 
study by \cite{sikora01} (see also \cite{moderski02}), who 
included a significant contribution of external Compton radiation 
to the high-energy emission of blazars, and focused on the modeling 
of photon-energy dependent light curves and time lags between 
different frequency bands. They applied their results to the
FSRQ 3C~279, and concluded that the correlated X-ray/$\gamma$-ray
variability of this quasar was inconsistent with X-rays and
$\gamma$-rays being produced by the same radiation mechanism 
because otherwise significant systematic time lags between the 
$\gamma$-ray and X-ray flaring behaviour would be expected,
contrary to the observations (e.g., \cite{hartman01}).

In the present paper, we describe a newly developed combined
SSC + ERC jet radiation transfer code, accounting for time-dependent
particle acceleration and injection, radiative cooling, and escape, 
coupled to the self-consistent treatment of the relevant photon
emission, absorption, and escape processes. In \S \ref{model} we
give a brief description of the underlying blazar jet model. The
numerical procedure used in our code will be outlined in \S \ref{numerics}.
We present results of a detailed parameter study, relevant for
application to intermediate and low-frequency peaked BL~Lac objects,
in \S \ref{results}. We summarize in \S \ref{summary}.

\section{\label{model}Model Description}

The blazar model used for this study is a generic leptonic jet model. 
It is assumed that a population of ultrarelativistic, non-thermal 
electrons (and positrons) is injected at a generally time-dependent 
rate into a spherical emitting volume of co-moving radius $R_b$ (the
``blob''). The injected pair population is specified through an injection 
power $L_{\rm inj} (t)$ and the spectral characteristics of the 
injected non-thermal electron distribution. We assume that electrons are
injected with a single power-law distribution with low and high energy
cutoffs $\gamma_1$ and $\gamma_2$, respectively, and a spectral index 
$q$ so that the injection function $Q_e (\gamma; t)$ [cm$^{-3}$~s$^{-1}$], 
in the co-moving frame of the emitting region, is 

\begin{equation}
Q_e^{\rm inj} (\gamma; t) = Q_0^{\rm inj} (t) \, \gamma^{-q} \; \; 
{\rm for} \;\; \gamma_1 \le \gamma \le \gamma_2
\label{Qe}
\end{equation}
with
\begin{equation}
Q_0^{\rm inj} (t) = \cases{ 
{L_{\rm inj} (t) \over V'_b \, m_e c^2} \, {2 - q \over \gamma_2^{2 - q} 
- \gamma_1^{2 - q}} & if $q \ne 2$ \cr\cr
{L_{\rm inj} (t) \over V'_b \, m_e c^2 \, \ln(\gamma_2/\gamma_1)} & if 
$q = 2$,\cr}
\label{Q0}
\end{equation}
where $V'_b$ is the blob volume in the co-moving frame.

The jet is powered by accretion of material onto a supermassive central
object, which is accompanied by the formation of an accretion disk. For the 
purpose of this study, we have represented the disk by a standard Shakura-Sunyaev
disk with a bolometric luminosity of $L_D = 10^{45}$~ergs~s$^{-1}$. 
The choice of this and several other standard parameters is motivated 
by a recent modeling study of the LBL W~Comae \citep{bmr02}. 
The randomly oriented magnetic field $B$ is determined by 
an equipartition parameter $\epsilon_B$, which is the fraction
of the magnetic field energy density $u_B$ compared to its value for
equipartition with the relativistic electron population in the emission
region. As a consequence of this parametrization, the  magnetic field 
will gradually change throughout the evolution of the blob as particles
are being injected and subsequently cool along the jet. The blob moves 
with relativistic speed $v / c = \beta_{\Gamma} = \sqrt{1 - 1 / \Gamma^2}$ 
along the jet which is directed at an angle $\theta_{\rm obs}$ (with 
$\mu \equiv \cos\theta_{\rm obs}$) with respect to the line of sight. 
The Doppler boosting of emission from the co-moving to the observer's 
frame is determined by the Doppler factor $D = \left[ \Gamma \, 
(1 - \beta\mu) \right]^{-1}$.

As the emission region moves outward along the jet, particles are 
continuously injected according to Eq. \ref{Qe}, are cooling, 
primarily due to radiative losses, and may leak out of the system. We 
parametrize particle escape through an energy-independent escape time 
scale $t_{\rm esc} = \eta \, R_b / c$ with $\eta \ge 1$. This 
parametrization of particle escape can also be used to include adiabatic 
losses, which are not explicitly taken into account in our simulations.
Radiation mechanisms included in our simulations are synchrotron emission, 
Compton upscattering of synchrotron photons (SSC = Synchrotron Self 
Compton scattering), and Compton upscattering of external photons 
(EC = External Compton scattering), including photons coming directly 
from the disk as well as re-processed photons from the broad line region. 
The broad line region is modelled as a spherical shell between $r_{\rm BLR, in} 
= 0.2$~pc and $r_{\rm BLR, out} = 0.25$~pc (we note, however, that 
the details of the radial distribution of the BLR material do not play an 
important role as long as $r_{\rm BLR, out} - r_{\rm BLR, in} \lesssim 
r_{\rm BLR, in}$), and a radial Thomson depth $\tau_{\rm T, BLR}$ which is 
considered a free parameter. $\gamma\gamma$ absorption and the corresponding 
pair production rates are taken into account self-consistently, using 
the general solution for the pair production rate of \cite{bs97}. However, 
in all simulations presented in this paper, the $\gamma\gamma$ opacity 
is $\ll 1$ out to at least several tens of GeV so that $\gamma\gamma$ 
absorption and pair production do not play an important role. Motivated 
by the $\sim 10$~hr minimum variability time scale observed in W~Comae 
\citep{tagliaferri00}, we choose $R_b = 10^{16}$~cm, and $\Gamma = D = 10$, 
which implies $\theta_{\rm obs} = 5.74^o$. 

Based on an equipartition parameter $\epsilon_B \sim 1$, we expect typical
magnetic field values of order $B \sim 1$~G \citep{bmr02}, which implies a 
synchrotron cooling time scale (in the observer's frame) of electrons emitting 
synchrotron radiation at an observed energy $E_{\rm sy} = 1 \, E_{\rm keV}$~keV
of

\begin{equation}
\tau_{\rm sy} \approx 0.29 \, \left( {B \over 1 \, {\rm G}} \right)^{-3/2} \,
\left( {D \over 10} \right)^{-1/2} \, E_{\rm keV}^{-1/2} \; {\rm hr}.
\label{tau_sy}
\end{equation}
For X-ray photon energies, this is shorter than the dynamical time scale 
$R_B / (D \, c)$, in agreement with the approximately symmetric shape of the 
X-ray light curves generally observed both in W~Comae \citep{tagliaferri00}
and BL~Lacertae \citep{ravasio02}. 

\section{\label{numerics}Numerical Procedure}

In order to treat the time-dependent electron dynamics and radiation
transfer problem in the emitting volume, we solve simultaneously the
kinetic equation for the relativistic electrons,

\begin{equation}
{\partial n_e (\gamma, t) \over \partial t} = - {\partial 
\over \partial \gamma} \left( \left[ {d\gamma \over dt} 
\right]_{\rm loss} \, n_e [\gamma, t] \right) + Q_e (\gamma, t) 
- {n_{\rm e} (\gamma, t) \over t_{\rm e, esc}},
\label{e_evolution}
\end{equation}
and for the photons,
\begin{equation}
{\partial n_{\rm ph} (\epsilon, t) \over \partial t} = 
\dot n_{\rm ph, em} (\epsilon, t) - \dot n_{\rm ph, abs}
(\epsilon, t) - {n_{\rm ph} (\epsilon, t) \over t_{\rm ph, esc}}.
\label{ph_evolution}
\end{equation}
Here, $(d\gamma / dt)_{\rm loss}$ is the radiative energy loss rate
for the electrons, $Q_e (\gamma, t)$ is the sum of the external 
injection rate $Q_e^{\rm inj}$ from Eq. \ref{Qe} and the intrinsic 
$\gamma\gamma$ pair production rate, $\dot n_{\rm ph, em} (\epsilon, t)$
and $\dot n_{\rm ph, abs} (\epsilon, t)$ are the photon emission and 
absorption rates corresponding to the various radiation mechanisms,
and $t_{\rm ph, esc} = (3/4) \, R_b/c$. In Eq. \ref{e_evolution}, 
electron cooling is approximated as a continuous function of time
(i.e., the energy of an individual electron is described as a 
differentiable function of time). This would be inaccurate if a 
significant contribution to the cooling rate were due to Compton 
scattering in the Klein-Nishina limit since in that case, the 
electron is transferring virtually all of its energy to a soft 
photon in a single scattering event. However, in the parameter 
ranges which we are primarily interested in, electron cooling 
is dominated by synchrotron losses and Compton scattering in 
the Thomson regime, for which Eq. \ref{e_evolution} is a good 
approximation. In Eq. \ref{ph_evolution}, the emissivity term
contains the contribution from Compton scattering into a given photon
energy interval. Since in all model situations considered here, the
Thomson depth of the emitting region is $\tau_{\rm T} \ll 10^{-6}$, 
the modification of the photon spectrum due to scattering of photons
out of a given photon energy range is negligible.  

The relevant electron cooling rates and photon emissivities and 
opacities are evaluated using the well-tested subroutines of the 
jet radiation transfer code described in detail in \cite{bms97} 
and \cite{bb00}. The full Klein-Nishina cross section for Compton 
scattering and the complete, analytical solution for the 
$\gamma\gamma$ pair production spectrum of \cite{bs97} are 
used. The discretized electron continuity equation can be
written in the form of a tri-diagonal matrix as in \cite{cg99},
which can be readily solved using the standard routine of 
\cite{press92}. This procedure turns out to be very stable
if, instead of the sharp cutoffs of the electron injection
function (\ref{Qe}), we introduce continuous transitions to very
steep power-laws to mimic these cutoffs. Specifically, we add
a low-energy branch with $Q_e (\gamma; t) \propto \gamma^2$
for $\gamma < \gamma_1$, and $Q_e (\gamma; t) \propto \gamma^{-(q+3)}$
for $\gamma > \gamma_2$. After each electron time step, we update 
the photon distribution using a simple explicit forward integration
of the discretized Eq. \ref{ph_evolution}. 

We have carefully tested our code by running it with parameters
identical to those used for Figs. 6 -- 14 of \cite{lk00}, who are
using a very similar numerical approach. We find very good agreement 
with their results, with only minor discrepancies which are due to
our replacing the high-and low-energy cutoff of the electron injection
spectrum by continuous transitions to very steep power-laws as described
above. Specifically, this results in more moderate spectral indices of 
the synchrotron spectra just beyond the high-energy cutoffs. 

\section{\label{results}Numerical Results}

We have performed a large number of simulations, studying the
influence of various model parameters on the resulting broadband
spectra, light curves, and X-ray hardness-intensity diagram tracks.
In each one of our simulations, we have assumed an underlying,
quiescent injection power of $L_{\rm inj}^{\rm qu} = 
10^{38}$~ergs~s$^{-1}$, on top of which we inject particles 
with various flaring injection powers in the range
$10^{40}$~ergs~s$^{-1} \; \le L_{\rm inj}^{\rm fl} \le
10^{43}$~ergs~s$^{-1}$. As a standard model setup, we choose
an injection electron function given by $\gamma_1 = 10^3$,
$\gamma_2 = 10^5$, and $q = 2.5$. In the base model, we
have $L_{\rm inj}^{\rm fl} = 10^{41}$~ergs~s$^{-1}$, 
extending as a step function in time over 2 dynamical
time scales, $\Delta t'_{\rm inj} = 2 R_b / c$ in the
co-moving frame. The injection event is centered around
a distance $x_0 = 0.1$~pc from the central engine. The
BLR Thomson depth is chosen to be 0 in the base model. 

Subsequently, we investigate the influence of changing
(1) the flaring injection power, (2) the BLR Thomson
depth and, accordingly, the contribution of external photons
to the soft photon field for Compton scattering, (3) the
electron injection spectral index $q$, (4) the duration of
the flaring injection event, (5) the time profile of the
flaring injection event, (6) the electron escape time scale
parameter. The parameters used for the individual runs are
quoted in Table \ref{parameters}. 

\begin{figure}
\plotone{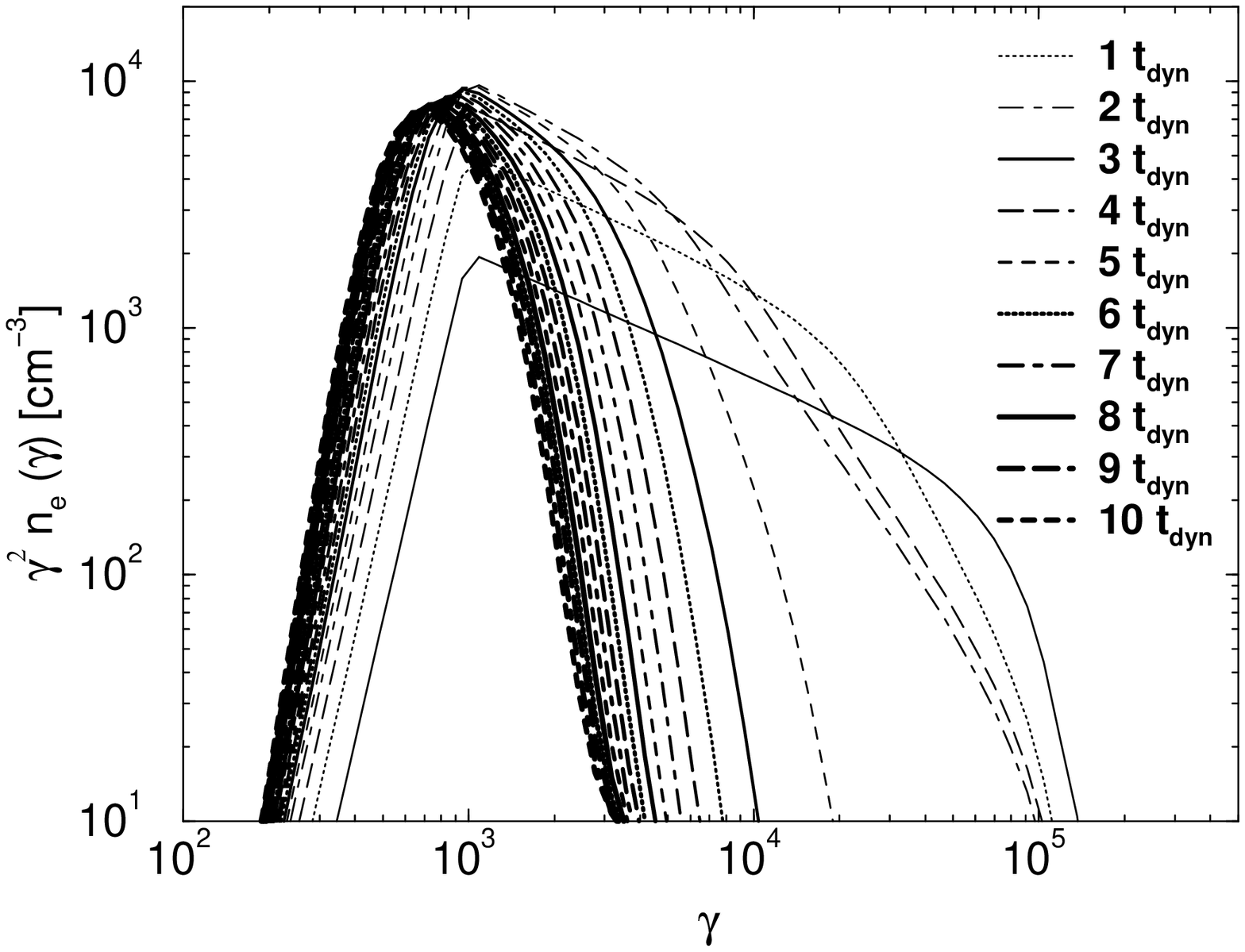}
\caption{Sequence of co-moving electron spectra in the emission region for 
our base model, simulation no. 2. The curves are labeled by time in multiples
of the dynamical time scale, $t_{\rm dyn}^{\rm obs} = R_b/(D \, c) =
3.33 \times 10^4$~s. For model parameters, see Tab. \ref{parameters}.}
\label{sim2_e}
\end{figure}

\begin{figure}
\plotone{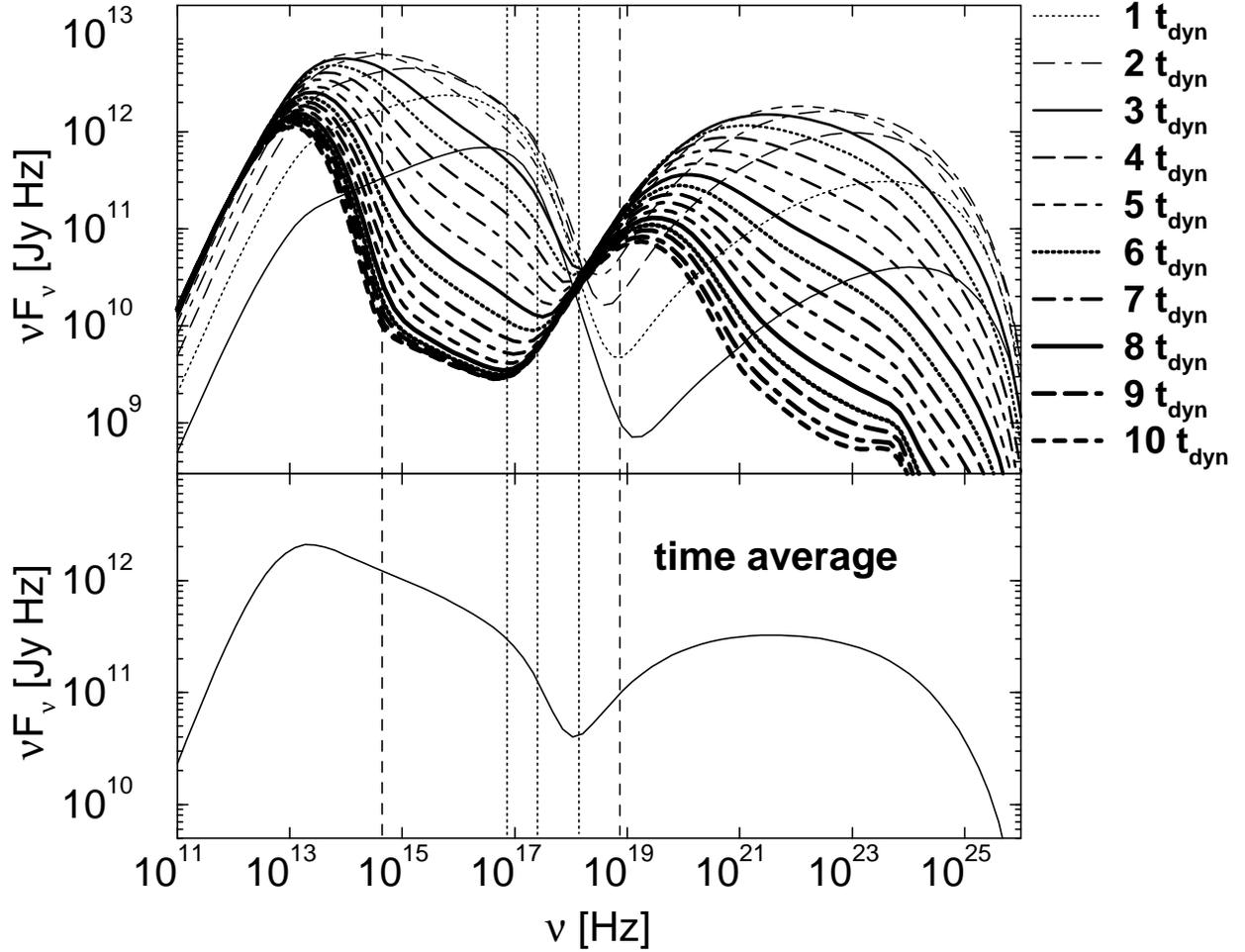}
\caption{Sequence of spectral energy distributions (upper panel) and the 
time averaged SED (lower panel) from our base model, simulation no. 2.
The dotted vertical lines indicate the frequencies at which light curves
and hardness-intensity correlations have been extracted; the dashed
vertical lines indicate the remaining two frequencies at which light
curves have been extracted.}
\label{sim2_ph}
\end{figure}

\begin{figure}
\plotone{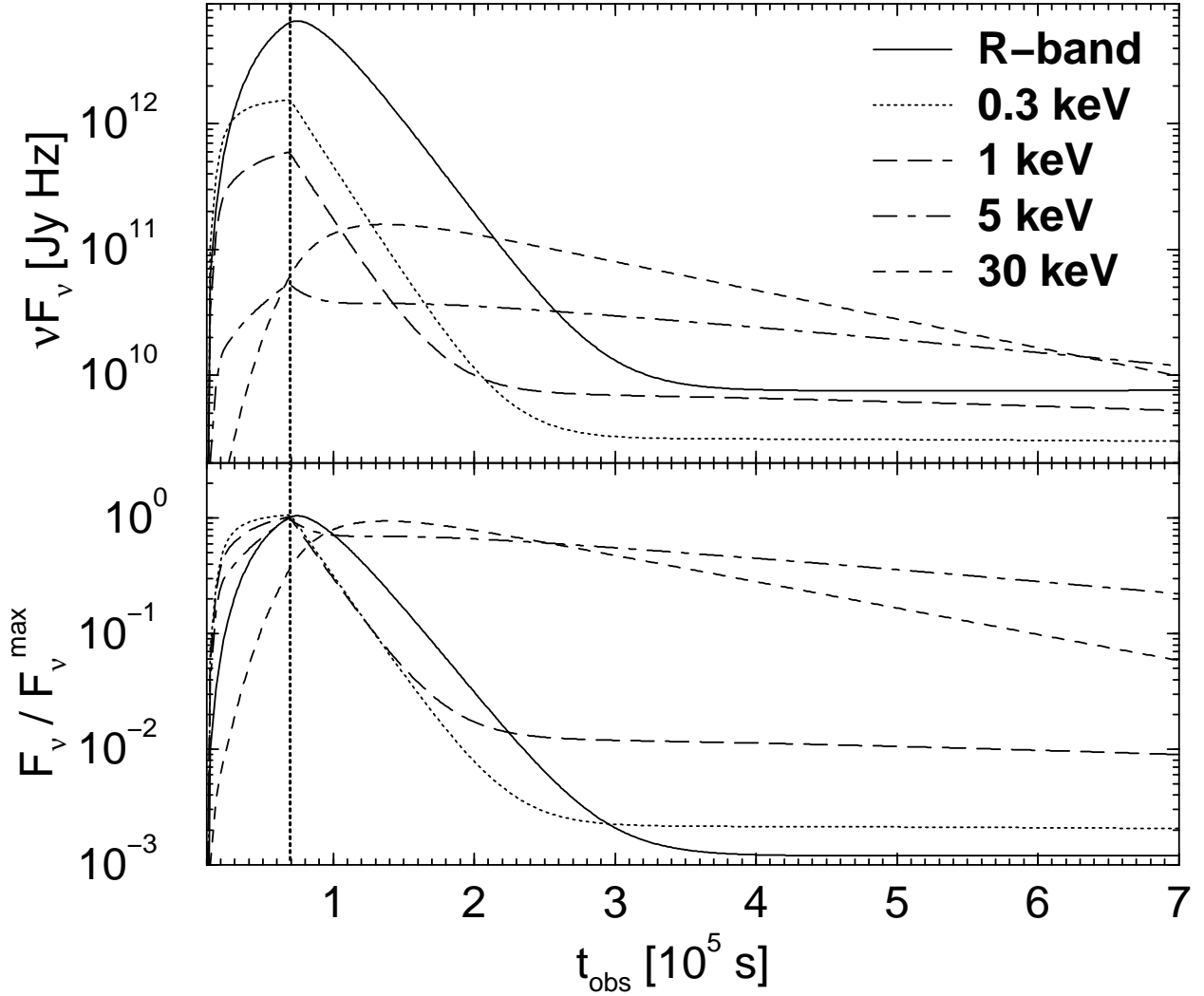}
\caption{Light curves at three different X-ray energies, in the optical
(R-band) and at hard X-rays for our base model. Upper panel: Absolute flux
values; lower panel: fluxes normalized to their peak values. The vertical 
dotted line indicates the end of the flaring injection episode.}
\label{sim2_lc}
\end{figure}

\begin{figure}
\plotone{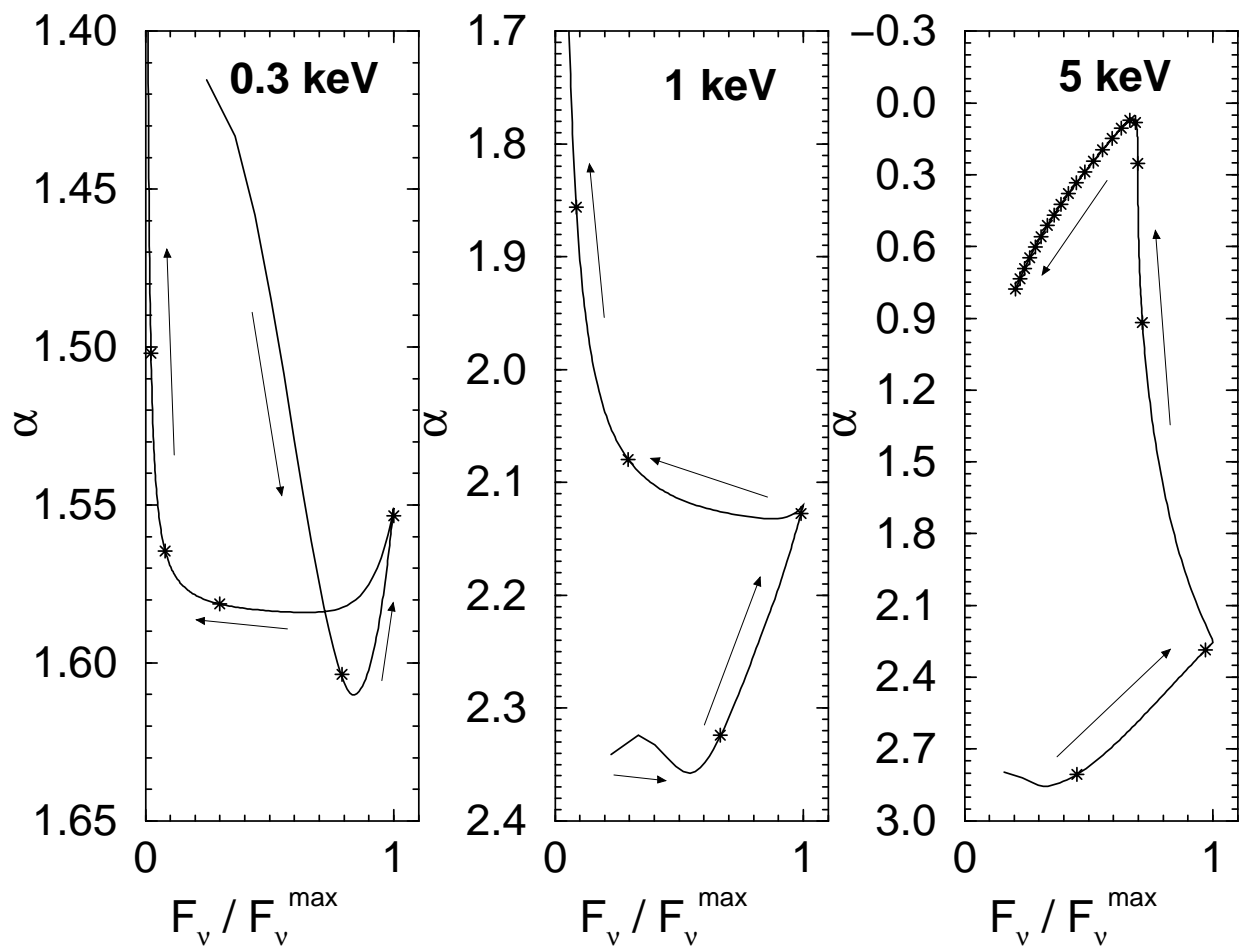}
\caption{Tracks of the simulated spectra in the hardness-intensity diagrams 
at three different X-ray energies for our base model. $\alpha$ is the local
energy spectral index $F_{\nu} \propto \nu^{- \alpha}$ at the respective
photon energy. Stars indicate the locations at multiples of the dynamical 
time scale during the simulation.}
\label{sim2_hic}
\end{figure}

Our base model is simulation no. 2. In Figs. \ref{sim2_e} and 
\ref{sim2_ph}, we have compiled a sequence of co-moving electron 
spectra, snap-shot SEDs, and the time averaged photon spectrum 
resulting from our base model. Light curves at 3 selected X-ray 
energies as well as in the optical (R-band) and at hard X-rays (30~keV)
are plotted in Fig. \ref{sim2_lc}, and tracks in the hardness-intensity 
diagrams (HIDs) at three different X-ray energies are compiled 
in Fig. \ref{sim2_hic}. The figures illustrate the gradual build-up
of the electron density in the emission region, competing with
radiative cooling, which is faster than the injection time scale
at electron energies of $\gamma \gtrsim 10^4$. Radiative cooling
is the dominant process affecting the electron spectra after
the end of the flaring injection episode at $t = 2 \, t_{\rm dyn}$.
The time-dependent photon spectra as well as the light curves
demonstrate that we do not expect significant peak time delays 
within the synchrotron component at frequencies $\nu \gtrsim 
10^{14}$~Hz, but that the high-energy (SSC) component is 
delayed by $\sim 1$ dynamical time scale due to the gradual 
accumulation of seed photons for Compton scattering. The 
figure also indicates the very moderate flux variability at
energies just above the synchrotron cut-off, which is located
at $\sim 1$~keV in our example. Fig. \ref{sim2_hic} illustrates
the spectral hysteresis phenomenon (keeping in mind that additional
contributions from previous injection episodes should close the
tracks in the sense that they are expected to start out near the
end points of the tracks shown in the figure). In agreement with
\cite{lk00} we find that --- at least in this generic case ---
the spectral hysteresis tracks can change their orientation
from clockwise to counterclockwise as one goes from photon
energies below the synchrotron cutoff to energies above the
cutoff, where the spectrum is dominated by Compton scattering 
(SSC). 

In the following, we are focusing on the time-averaged photon
spectra, the light curves, and the X-ray spectral hysteresis, 
and investigate how those aspects are affected by variations 
of individual parameters. 

\subsection{\label{power}Electron Injection Power}

The effect of an increasing injection power --- corresponding to a 
higher density of injected, relativistic particles in the emitting 
region --- is illustrated in Figs. \ref{L_inj_intspectra} -- 
\ref{L_inj_hic}. In addition to a corresponding increase in the
overall bolometric luminosity, this leads also to a stronger
relative energy output in the SSC-dominated Compton emission
at X- and $\gamma$-ray energies, as expressed, e.g., in 
Eq. (19) of \cite{cb02}. The photon spectral index of
the time-averaged emission at optical -- soft X-ray frequencies 
remains robust at $\alpha_{o - X} \sim 1.25$ due to optically
thin synchrotron emission from the cooled electron spectrum
with injection spectral index $q = 2.5$. As the bolometric
luminosity (and the electron cooling) becomes dominated by
the SSC mechanism, one would expect that this changes to the 
canonical value of $\alpha_{o - X} = 1.5$, which is a result 
of the decaying electron cooling rate in an SSC-dominated cooling 
scenario \citep{cb02}. However, in the situation simulated 
here, the first-order SSC peak is rapidly (within $\sim 3 \, 
t_{\rm dyn}$) decaying into the keV energy range and dominating
over the instantaneous synchrotron emission at UV -- X-ray energies. 
This leads to a significant hardening of the time-averaged optical 
-- X-ray spectrum, which even becomes inverted in $\nu F_{\nu}$
space in our most extreme test case (simulation no. 8).

In the light curves (see Fig. \ref{L_inj_lc}), the more rapid
electron cooling with increasing electron injection power manifests 
itself in an overall increasing amplitude of variability at all
energies. In particular, as SSC cooling becomes more important,
even the harder X-rays begin to exhibit significant variability
on the dynamical time scale, in contrast to the synchrotron-cooling 
dominated cases. Furthermore, while for very low injection powers,
the synchrotron cooling time scale for optical synchrotron emission
is comparable to the injection time scale, resulting in a time delay 
of $\sim$~a few hr between X-ray and optical emission, the optical 
light curve peaks at the end of the injection episode for higher 
injection powers, simultaneously with the X-rays. However, when SSC
cooling becomes dominant, the gradually increasing energy density
in the soft photon field during the injection episode actually has
the effect that the X-ray light curves are peaking at the beginning
of the injection episode, which would, again, lead to a time delay
of $\sim $~a few hr between X-ray and optical flares. 

Fig. \ref{L_inj_hic} illustrates how the tracks in the HIDs at
different X-ray energies are drastically changing for different
injection powers. In particular at X-ray energies just below or
at the synchrotron cutoff ($\sim 1$~keV), the flux maxima are
occurring at significantly different values of the local spectral
index $\alpha$ for different values of the injection power. 
Specifically, the local spectral indices at the time of the 
peak flux are significantly smaller (harder) for larger values 
of the injection power. Obvious changes in the orientation of 
the spectral hysteresis tracks are not found in these simulations. 

As mentioned earlier (see Eq. \ref{tau_sy}), in the test cases 
investigated here, the synchrotron cooling time scale of electrons 
emitting synchrotron radiation at X-ray energies, is shorter than 
the dynamical time scale, which is of the same order as the injection
time scale. Consequently, our results can be qualitatively compared 
to those of \cite{lk00} for the ``short cooling time limit'', bearing
in mind that the parameter values in our simulations have been chosen
appropriate for intermediate and LBLs, while \cite{lk00} focused on
the application to the HBL Mrk~421. In particular the light curves
displayed in their Figs. 9 -- 11 exhibit the same general trends as
we have found in our set of simulations. 

\subsection{\label{external}External Photons}

In order to investigate the influence of an increasing contribution
of external photons to the soft photon field for Compton scattering,
we performed a series of simulations with increasing values of 
$\tau_{\rm T, BLR}$, from 0 to 1. We note that the contribution 
from direct accretion disk photons to the photon energy density in
the emitting region is negligible in our base model. Thus, in the 
following, the external photons are primarily accretion disk photons
reprocessed in the BLR. An additional component due to direct accretion
disk photons would become energetically important only for injection
flares significantly contributing at $x_0^{\rm ECD} \lesssim 10^{-2}
\, L_{45}^{1/2} / [(B / {\rm G}) \, \Gamma_1]$~pc, where $L_{45}$ is 
the accretion disk luminosity in units of $10^{45}$~ergs~s$^{-1}$ and
$\Gamma_1 = \Gamma / 10$. For a recent discussion of this radiation 
field on the high-energy light curves see, e.g., \cite{ds02}.

The time-averaged photon spectra from our simulations with varying
BLR Thomson depth are shown in Fig. \ref{ext_intspectra}, which 
clearly shows the emergence of the external Compton (EC) component 
at GeV $\gamma$-ray energies. The impact of this additional emission 
component on the lower-frequency emission is small as long as its
bolometric energy output is smaller than or comparable to the 
synchrotron $\nu F_{\nu}$ flux. Only when EC cooling becomes
dominant over synchrotron cooling, are the effects on the 
synchrotron + SSC dominated portion of the spectrum (radio frequencies
-- MeV $\gamma$-rays) noticeable. This is the case when the comoving 
energy densities of the magnetic field and the external photons, 
$u'_{\rm B}$ and $u'_{\rm ext}$, respectively, become comparable. 
From

\begin{equation}
{u'_{\rm ext} \over u'_{\rm B}} \sim 20 { L_{45} \, \tau_{\rm T, BLR}
\, \Gamma_1^2 \over (B/{\rm G})^2 \, r_{0.2}^2}
\label{u_comparison}
\end{equation}
--- where $r_{0.2}$ is the radius of the inner boundary of the BLR in 
units of 0.2~pc --- we can estimate that this happens at $\tau_{\rm T, BLR} 
\sim 0.1$. Specifically, for $\tau_{\rm T, BLR} \gtrsim 0.1$, the effect 
of dominant external-Compton cooling results in a reduction of the 
time-averaged emission around the synchrotron peak --- leading to 
a spectral hardening of the synchrotron emission ---, and a suppression 
of the SSC emission. The suppression of the low-frequency synchrotron 
emission can be explained as the combined effect of two causes: First, 
electrons at energies above the low-energy cut-off are radiatively cooling 
on a time scale much shorter than the dynamical one (see Eq. \ref{B_decay} 
below). Consequently, the particle spectrum of high-energy electrons
injected during the flare is rapidly depleted within less than one 
dynamical time scale from the end of the flaring episode. However, 
after this episode, we are still injecting electrons (although at 
a much smaller rate corresponding to $L_{\rm inj}^{\rm qu}$) into
the blob, which continue to present a high-energy electron population.
In the case of the extremely fast cooling rate at $\tau_{\rm T, BLR}
\sim 1$, this additional high-energy electron population leads to a
significant additional contribution of synchrotron emission at intermediate
energies beyond the down-shifted maximum energy of electrons injected
during the flare and, consequently, to a hardening of the synchrotron
spectrum.

The second cause of spectral hardening of the low-frequency synchrotron
spectrum is related to our parametrization of the magnetic field 
in terms of an equipartition parameter $\epsilon_B$: For electron 
injection spectral indices $q > 2$, most of the energy in the electron 
population is carried by electrons near the low-energy cutoff. 
Consequently, the co-moving magnetic field will decay on a time 
scale given by the radiative cooling time scale of the lowest-energy
electrons, which is

\begin{equation}
t'_{\rm B} = t'_{\rm EC} (\gamma_1) \approx 4 \times 10^4 \, {r_{0.2}^2
\over \tau_{\rm T, BLR} \, \gamma_{1,3} \, L_{45} \, \Gamma_1^2} \; 
{\rm s}
\label{B_decay}
\end{equation}
where $\gamma_{1,3} = \gamma_1 / 10^3$. For $\tau_{\rm T, BLR} 
\gtrsim 0.1$, the time scale (\ref{B_decay}) is of the order 
of or shorter than the dynamical time scale. In contrast, the 
energy density of the external radiation field remains 
approximately constant during the evolution of the blob.
Consequently, as radiative cooling proceeds and shifts the electron
distribution towards lower energies, a steadily decreasing fraction
of the electron energy will be converted to synchrotron radiation. 
This leads to a spectral hardening of the time-averaged synchrotron 
spectrum compared to the canonical $\alpha = q/2$ spectrum above
the break frequency $\nu_1 = \nu_{L, 0} \, \gamma_1^2$ of a cooling
electron population in a constant magnetic field $B \equiv B_0$
with $\nu_{L, 0} = e B_0 / (2 \pi m_e c)$. At frequencies below 
$\nu_1$, the time-averaged spectrum would have a slope of $\alpha 
= 1/2$ in the constant-magnetic-field case. The steepening of this
low-frequency part of the synchrotron spectrum in the case of a
magnetic field proportional to the equipartition value can be derived
analytically in the following way.

At late times, flaring electron distribution has basically collapsed
to a $\delta$ function in electron energy, i. e. $n(\gamma_e, t) \propto
\delta(\gamma_e - \gamma[t])$. Then, the synchrotron emission coefficient
will behave as

\begin{eqnarray}
j_{\nu, \rm sy} (t) 
  &\propto& \dot{\gamma}_{\rm sy} \, \delta(\nu - \gamma[t]^2 \nu_L[t]) \\
  &\propto& \gamma[t]^3 \, \delta(\nu - \gamma[t]^{5/2} \nu_{L,0}).
\end{eqnarray}
where $\dot{\gamma}_{\rm sy} \propto \gamma(t)^3$ is the synchrotron loss 
rate. The additional factor of $\gamma$ results from the equipartition 
prescription, $B^2/8\pi = \epsilon_B n_e \gamma$. The electron evolution 
is governed by EC losses, and since the EC photon field is essentially 
constant over the flare episode, the electron Lorentz factor still 
evolves according to $\dot{\gamma} \approx -(4/3) (\sigma_T/m_e c) 
u_{\rm ext}^\prime \, \gamma^2$.  The time-integrated synchrotron 
spectrum is then
\begin{eqnarray}
\langle j_{\nu, \rm syn} \rangle_t 
       &\propto& \int dt \, \gamma[t]^3 \delta(\nu - \gamma[t]^{5/2}
        \nu_{L,0}) \\
       &\propto& \gamma[t]^3 \left|\frac{d\gamma[t]^{5/2}}{dt}\right|^{-1} 
        \propto \nu^{-1/5}. 
\end{eqnarray}
where we have used $\nu \propto \gamma^{5/2}$. The short-dashed curve
in Fig. \ref{ext_intspectra} shows the result of a test calculation in
which we held the magnetic field constant, while all other parameters
were identical to the $\tau_{\rm T, BLR} = 1$ simulation, in order to 
verify that the spectral hardening at radio frequencies is indeed 
partially a consequence of our magnetic field parametrization.

Fig. \ref{ext_lc} shows that the impact of a strong external Compton
component on the optical and X-ray light curves is very moderate. In
particular, the light curves at energies below the synchrotron cutoff
remain virtually unchanged. However, a strong external Compton component,
dominating the bolometric luminosity ($\tau_{\rm T, BLR} \gtrsim 0.1$
in our case) leads to a significantly faster decay of the light curves
at X-ray energies beyond the synchrotron cutoff.

Just as the light curves, also the X-ray spectral hysteresis characteristics
remain virtually unchanged, even in the case of a strongly dominant external
Compton component, except for a moderate softening of the local spectrum in
the decaying phase of the flare at energies beyond the synchrotron cutoff 
(see Fig. \ref{ext_hic}).

\subsection{\label{spectral_index}Electron Spectral Index}

The value of the electron injection spectral index $q$ should be
rather easily determined by measuring the time-averaged optical --
X-ray spectral index of the strongly cooled synchrotron spectrum. 
As illustrated in Fig. \ref{q_intspectra}, this spectral change is
accompanied by a shift of the SSC peak towards higher frequencies 
as the injection spectrum hardens: For $q > 2.5$, the SSC peak is
located at $\epsilon_{\rm SSC} \sim \gamma_1^2 \epsilon_{\rm sy}$,
while for $q < 2.5$, it shifts towards $\epsilon_{\rm SSC} \sim 
\gamma_2^2 \epsilon_{\rm sy}$. As illustrated in Fig. \ref{q_lc},
the characteristics of the light curves are only marginally different 
for different values of $q$. The X-ray spectral hysteresis at
X-ray energies below the synchrotron cutoff is obviously shifted
according to the change in injection spectral index, but its
basic characteristics remain unchanged (see Fig. \ref{q_hic}). 
An interesting qualitative change of the spectral hysteresis can
be seen for energies just above the synchrotron cutoff: While for 
hard electron injection spectra ($q < 2.5$), the peak flux is 
reached at a steep local X-ray spectrum (i.e. the spectrum is 
dominated by the synchrotron component at the time of peak flux),
a soft injection spectrum ($q > 2.5$) leads to a hard local
spectral index at the time of peak flux (i.e. the spectrum is
dominated by the SSC component at that time). 

\subsection{\label{duration}Duration of the Flare}

In order to investigate the influence of the duration of the
electron injection event causing the flare, we have performed
simulations with 3 different values of $t_{\rm inj}$, where
we kept the total energy input during the flare constant
(simulations no. 2, 11, and 12). The time-averaged SEDs from
those simulations are virtually indistinguishable. Fig. 
\ref{duration_lc} shows that the longer injection time scale 
at a lower injection power leads to a more gradual rise of 
the light curves at energies above the synchrotron cut-off, 
whereas at lower frequencies, this leads to a rather marginal 
modification of the rising portion of the light curve, followed 
by an extended plateau until the end of the injection episode. 
The decaying portion of the light curves at optical and X-ray 
frequencies seems to be virtually independent of the duration 
of the injection event. In the X-ray spectral hysteresis (Fig. 
\ref{duration_hic}), we find that for longer electron injection
events, the rising-flux portion of the track in the HID at
X-ray energies below the synchrotron cutoff occurs with increasingly
softer local spectra, whereas the decaying portion remains
virtually unchanged. At energies above the synchrotron peak,
the trend concerning the rising portion is opposite: With
longer duration of the injection episode, the local spectra
are becoming harder. 

From an observational point of view, one should be able 
to distinguish situations in which the duration of the 
flaring event is substantially longer than the dynamical
time scale, by virtue of the extended high-flux plateaus at
energies below the synchrotron peak, which are generally not 
observed in BL~Lac objects in which there is evidence for a
substantial contribution from synchrotron emission to the
X-ray flux (e.g., \cite{tagliaferri00,ravasio02}). This seems 
to indicate that the generic situation of $t_{\rm inj} \sim 2 
\, t_{\rm dyn}$ might be a realistic assumption.

\subsection{\label{timeprofile}Time Profile of the Flare}

The step function time profile of the electron injection power
during the flare is certainly a rather crude over-simplification 
of any realistic acceleration scenario. In order to investigate
whether this particular choice of the time profile has a significant
impact on our results, we have calculated an additional set of
simulations (nos. 13 -- 15), with triangular injection profiles.
Here, we have introduced a linear rise and decay of the injection
power on time scales $t_r$ and $t_d$, respectively, and have chosen
the maximum of the profile to be twice the injection power of the 
step-function case in order to keep the total injected energy at the 
same value as in our base model. 

The time-averaged photon spectra for all of these cases are virtually 
identical. Equally, the light curves at energies above the synchrotron
cutoff are only marginally affected by the details of the injection
time profile (see Fig. \ref{profile_lc}), while at lower energies the
light curves tend to track the injection time profile to a certain
extent during the rising portion of the light curves. The decaying
portion of the light curves is generally independent of the injection
time profile for all optical and X-ray photon energies. Fig. 
\ref{profile_hic} illustrates that the impact of the detailed injection 
profile on the X-ray spectral hysteresis characteristics is rather 
moderate, even for the extreme (and very artificial) cases of the
time profiles c (gradual rise over $2 \, t_{\rm dyn}$, sharp decay) 
and d (sharp rise, gradual decay over $2 \, t_{\rm dyn}$) illustrated 
here. Its influence is basically restricted to the flux-rise portion 
of the HID track and to photon energies below the synchrotron cut-off, 
where the local spectra tend to be harder for time profiles with maxima 
closer to the onset of the flare. The HID tracks for all our test cases 
show almost identical spectral indices at the time of maximum flux. 

\subsection{\label{escape}Electron Escape Time Scale}

At the low-energy end of the electron spectra, particles cooling
down from higher energies will either accumulate and build up a
$\gamma^{-2}$ power-law spectrum at energies below $\gamma_1$, or
escape, depending on the value of the escape time scale parameter
$\eta$. We can define a critical escape parameter 

\begin{equation}
\eta_{\rm cr} = {c \over R \, \tau_{\rm sy} (\gamma_1)}
= 2.3 \, \left( {R \over 10^{16} \; {\rm cm}} \right)^{-1} 
\left( {B \over {\rm G}} \right)^{-2} \, \left( {\gamma_1
\over 10^{3}} \right)^{-1}
\label{eta_cr}
\end{equation}
for which the escape time scale for electrons at energy $\gamma_1$ 
equals the synchrotron cooling time scale. For $\eta \le \eta_{\rm cr}$,
the electron distribution will maintain a sharp low-energy cutoff at 
$\gamma_1$, while for $\eta \gg \eta_{\rm cr}$, a -2 low-energy power
law will develop. In order to investigate whether our choice of 
$\eta = 10$ has a significant impact on our results, we have done
test simulations with $\eta = 3$ and $\eta = 30$, respectively.
We find all relevant aspects --- the time-averaged spectra, the
monoenergetic light curves, and the spectral hysteresis curves ---
to be virtually independent of $\eta$ within reasonable bounds.

\subsection{\label{other}Other parameters}

In the previous subsections, we have discussed the impact of various
parameters on the broadband SED and the X-ray variability characteristics 
of models for intermediate and low-frequency peaked BL~Lac objects. 
There are still a few more parameters left which we have fixed in
our suite of test simulations. In particular, the choice of the
cutoffs of the electron distribution, $\gamma_1$ and $\gamma_2$,
the magnetic-field equipartition parameter $\epsilon_B$, the Doppler
factor $D$, and the size of the emitting region, $R_b$, may have an 
impact on the spectral variability characteristics. However, those 
parameters can generally be constrained rather well through the
observed overall spectral characteristics, and through variability
time scale considerations (see, e.g., \cite{tmg98,cg02}). The values 
adopted here are representative for typical LBLs like BL~Lacertae 
\citep{madejski99,bb00} or W~Comae \citep{tagliaferri00,bmr02}.
Furthermore, significantly different values of $\gamma_2$, $\epsilon_B$,
and $D$ would shift the synchrotron cutoff out of the X-ray regime 
so that the diagnostics developed here may not be applicable to the
X-ray variability of BL~Lac objects, which is the focus of this
paper. 

\section{\label{summary}Summary and Conclusions}

We have presented a detailed parameter study of the time-dependent
electron injection and kinematics and the self-consistent radiation
transport in jets of intermediate and low-frequency peaked BL~Lac
objects. Those objects are currently of great interest as the
steadily improving capabilities of current and future air \v Cerenkov
detector facilities might allow the detection of this class of blazars
at multi-GeV energies in the near future. At the same time, some of
these objects exhibit interesting X-ray variability features which
can now be studied in detail with the new generation of X-ray
telescopes, in particular {\it Chandra} and {\it XMM-Newton}.
Furthermore, the GLAST mission, scheduled for launch in 2006, is
expected to detect many more BL~Lac objects at multi-MeV -- GeV
energies and bridge the observational gap between the energy
ranges previously covered by the EGRET instrument on board 
{\it CGRO}, and the ground-based air \v Cerenkov facilities.

In our study, we have focused on the impact of various specific
parameter choices and variations on the broadband SEDs, optical
and X-ray light curves, and the spectral hysteresis phenomena
previously observed in several high-frequency peaked BL~Lac
objects, but also expected to be observable in intermediate
and low-frequency peaked BL~Lacs. 

\begin{figure}
\plotone{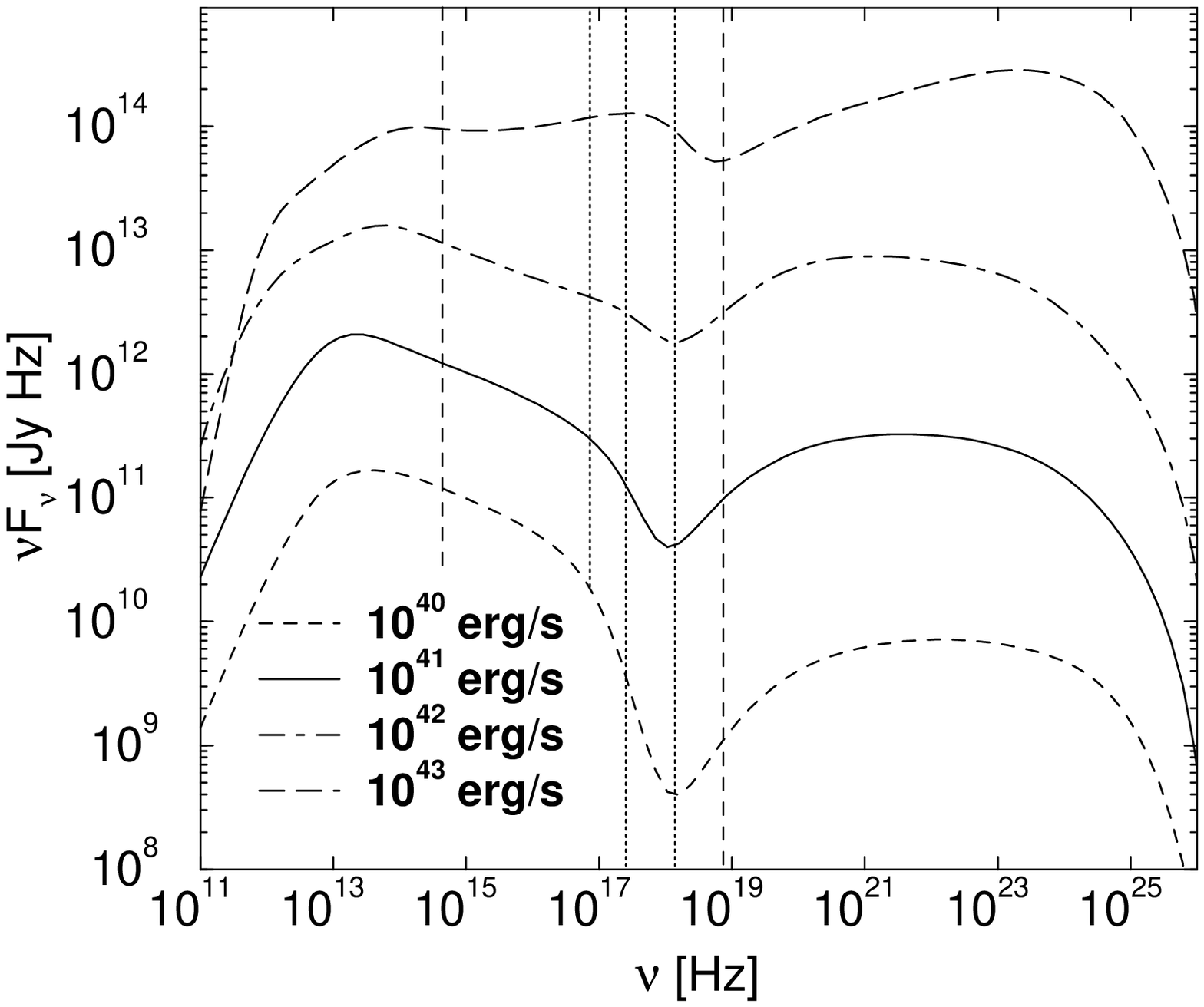}
\caption{Time averaged photon spectra for different values of the flaring
injection power $L_{\rm inj}^{\rm fl}$, from simulations no. 1, 2, 3, and 8.
The dotted vertical lines indicate the frequencies at which light curves
and hardness-intensity correlations have been extracted; the dashed
vertical lines indicate the remaining two frequencies at which light
curves have been extracted.}
\label{L_inj_intspectra}
\end{figure}

\begin{figure}
\plotone{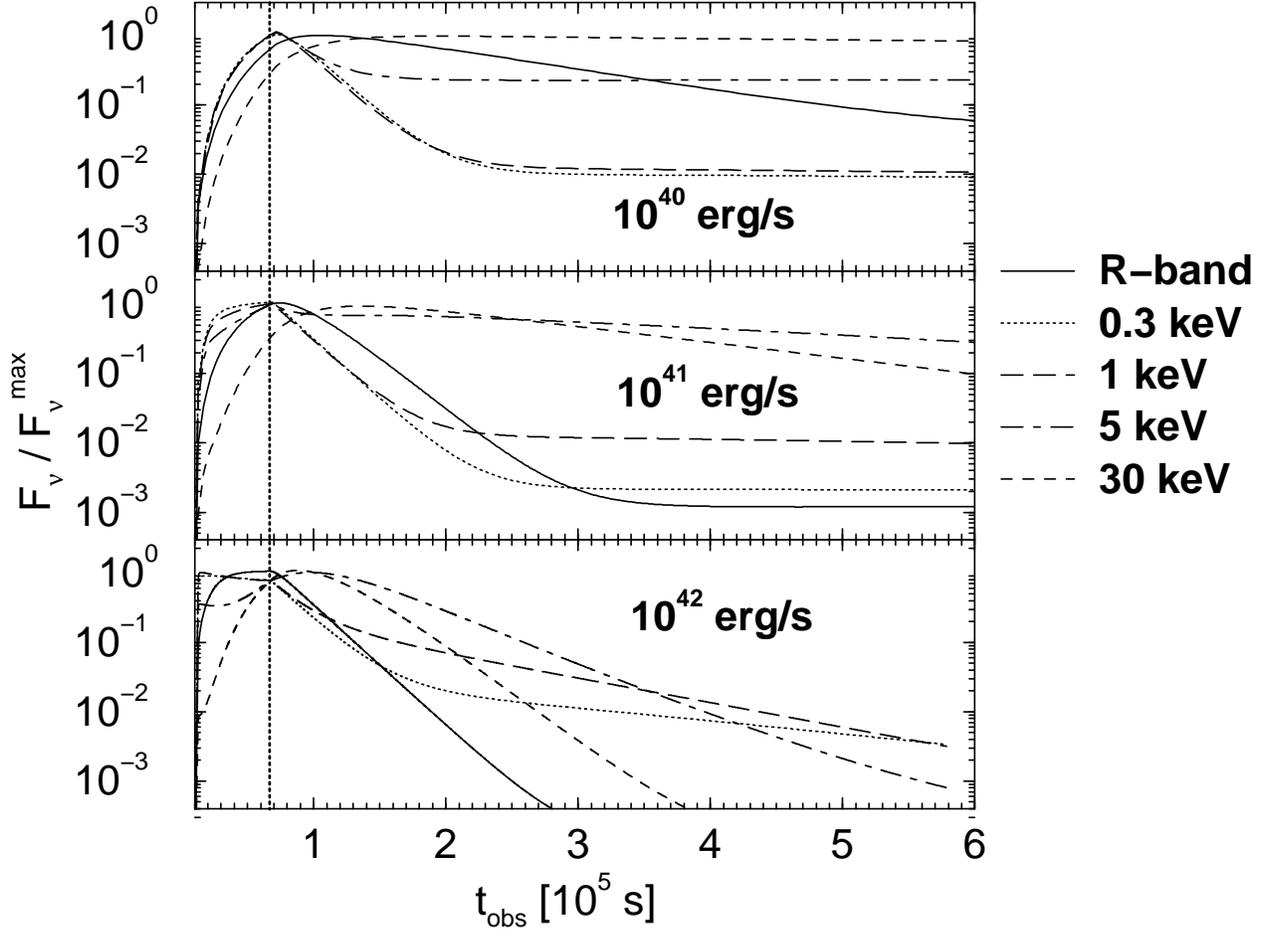}
\caption{Optical and X-ray light curves for different values of the flaring
injection power $L_{\rm inj}^{\rm fl}$, from simulations no. 1, 2, and 3.}
\label{L_inj_lc}
\end{figure}

\begin{figure}
\plotone{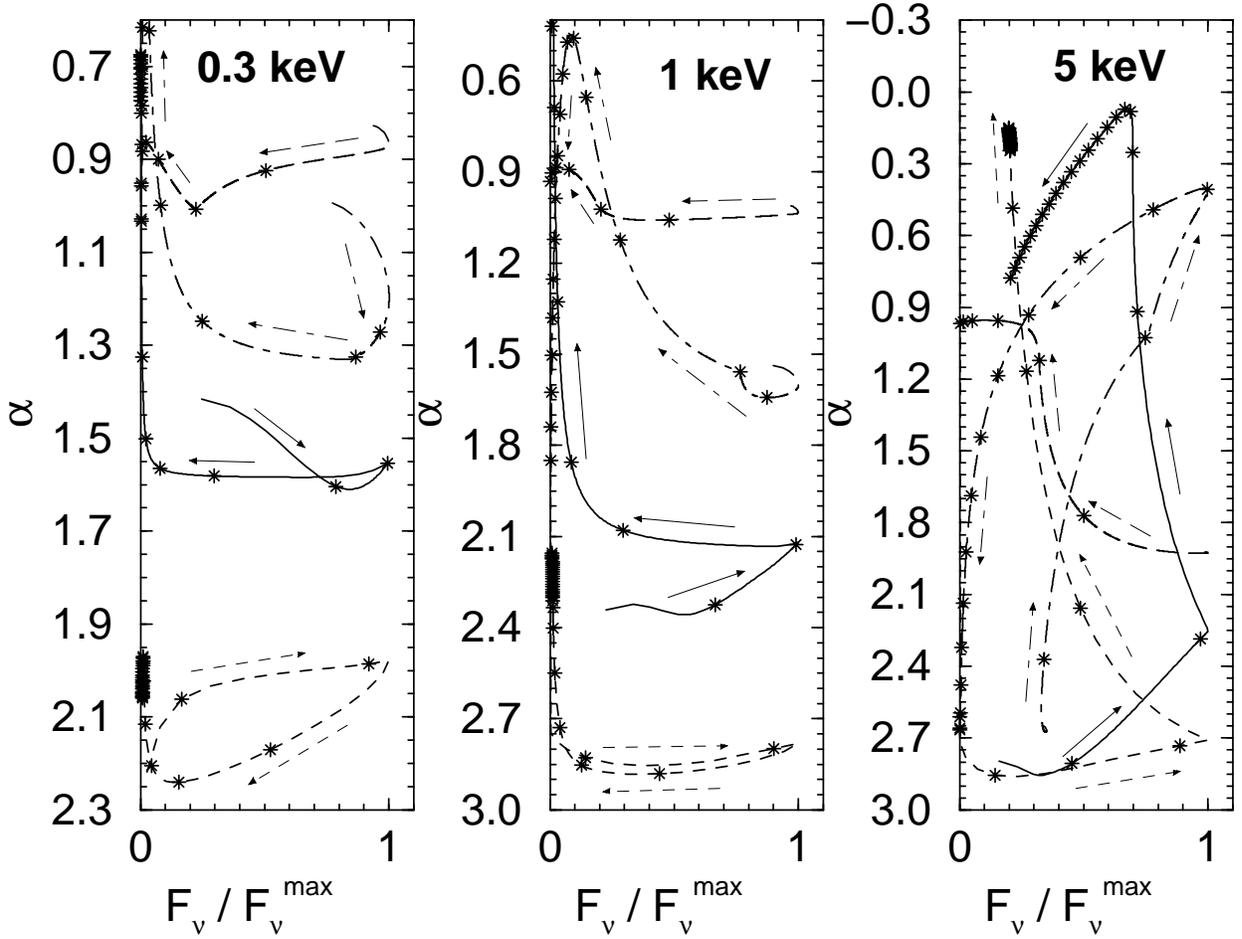}
\caption{Tracks in the harness-intensity diagrams at X-ray energies for
different values of the flaring injection power $L_{\rm inj}^{\rm fl}$, 
from simulations no. 1, 2, 3, and 8. Symbols are the same as in 
Fig. \ref{L_inj_intspectra}: short-dashed ($10^{40}$~erg/s), solid
($10^{41}$~erg/s), dot-dashed ($10^{42}$~erg/s), long-dashed 
($10^{43}$~erg/s). Stars indicate the locations at multiples of the 
dynamical time scale during the respective simulations.}
\label{L_inj_hic}
\end{figure}

\begin{figure}
\plotone{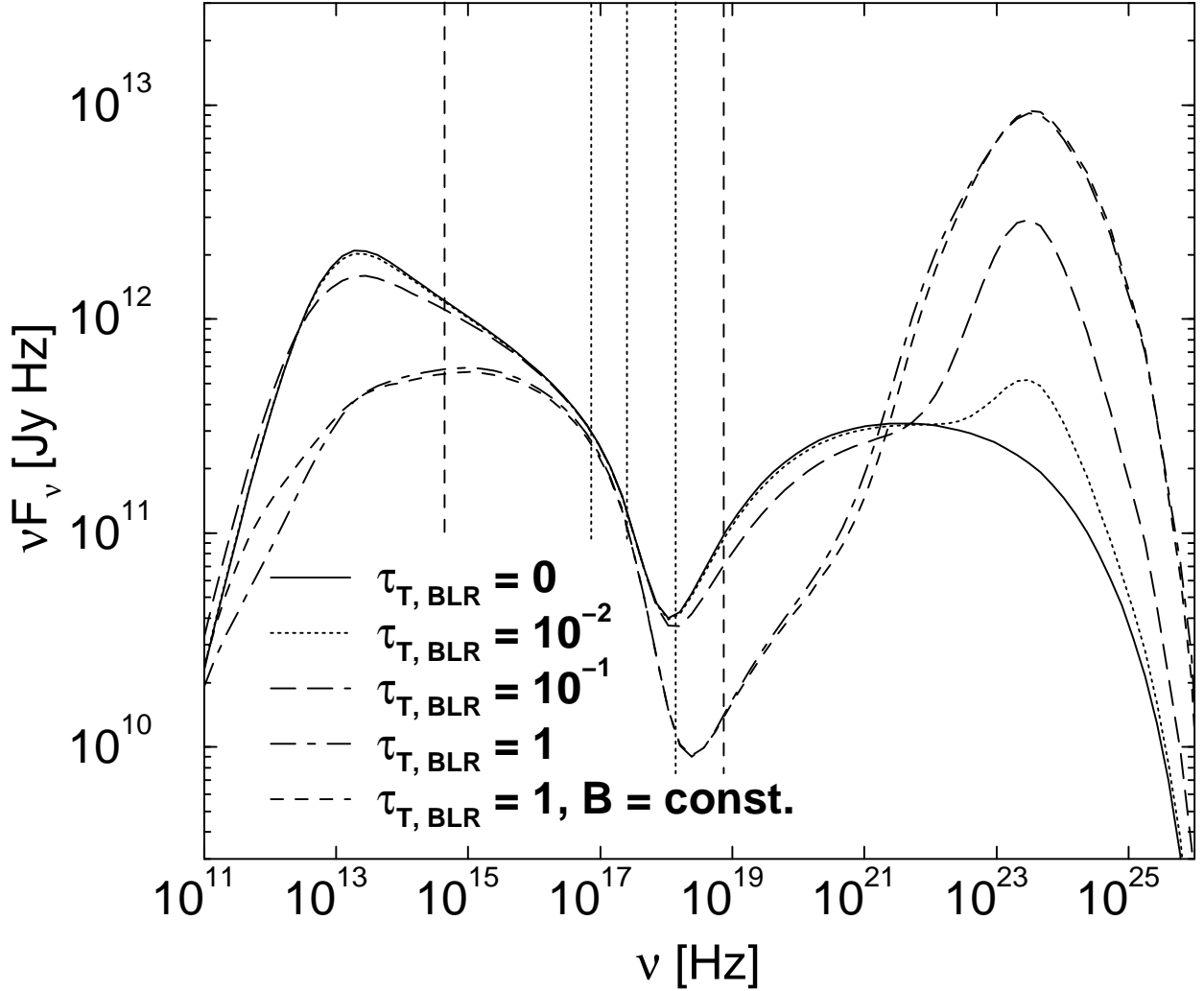}
\caption{Time averaged photon spectra for different intensities of the
external soft radiation field, parametrized through different values of
$\tau_{\rm T, BLR}$, from simulations no. 2, 5, 6, 7, and 18. The 
short-dashed curve shows a test simulation with constant magnetic 
field, while all other parameters were identical to the $\tau_{\rm T, 
BLR} = 1$ simulation, illustrating the effect of our magnetic field 
parametrization as a constant fraction of the equipartition magnetic
field. The dotted vertical lines indicate the frequencies at which 
light curves and hardness-intensity correlations have been extracted; 
the dashed vertical lines indicate the remaining two frequencies at 
which light curves have been extracted.}
\label{ext_intspectra}
\end{figure}

\begin{figure}
\plotone{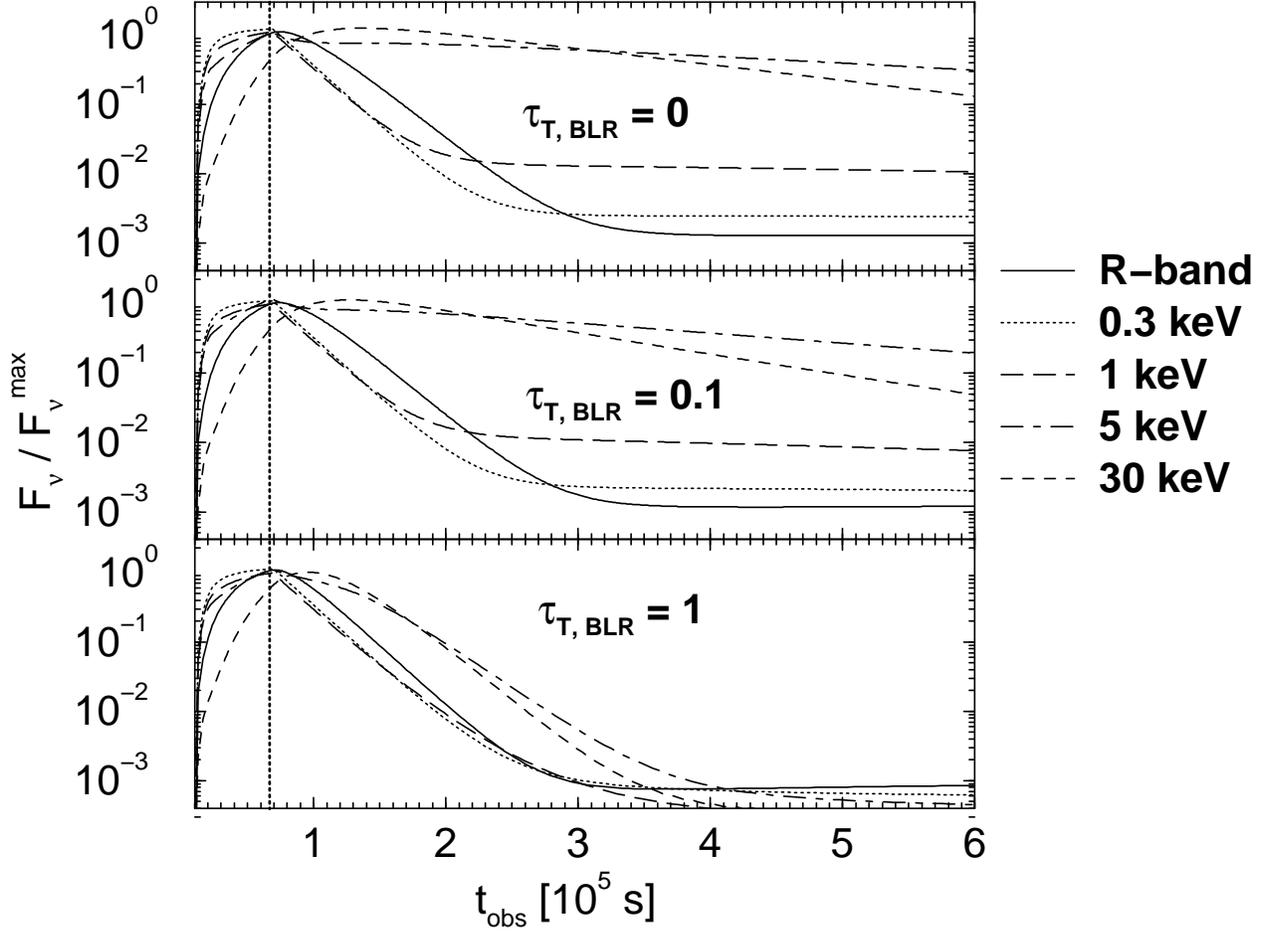}
\caption{Optical and X-ray light curves for different intensities of the
external soft radiation field, parametrized through different values of
$\tau_{\rm T, BLR}$, from simulations no. 2, 6, and 7.}
\label{ext_lc}
\end{figure}

\begin{figure}
\plotone{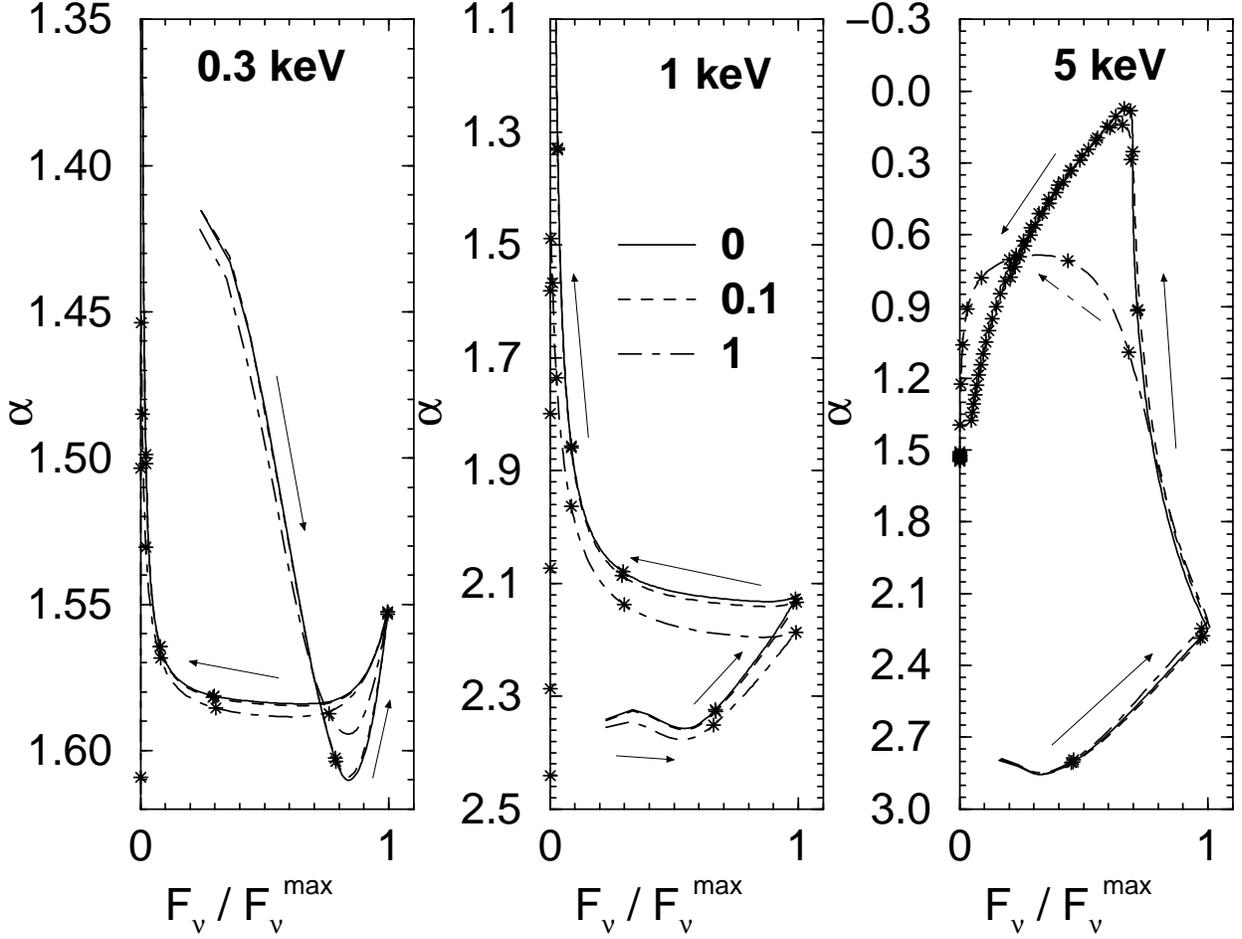}
\caption{Tracks in the harness-intensity diagrams at X-ray energies for
different intensities of the external soft radiation field, parametrized 
through different values of $\tau_{\rm T, BLR}$, from simulations no. 2, 6, 
and 7: solid ($\tau_{\rm T, BLR} = 0$), dashed ($\tau_{\rm T, BLR} = 0.1$),
dot-dashed ($\tau_{\rm T, BLR} = 1$). Stars indicate the locations at
multiples of the dynamical time scale during the respective simulations.}
\label{ext_hic}
\end{figure}

\begin{figure}
\plotone{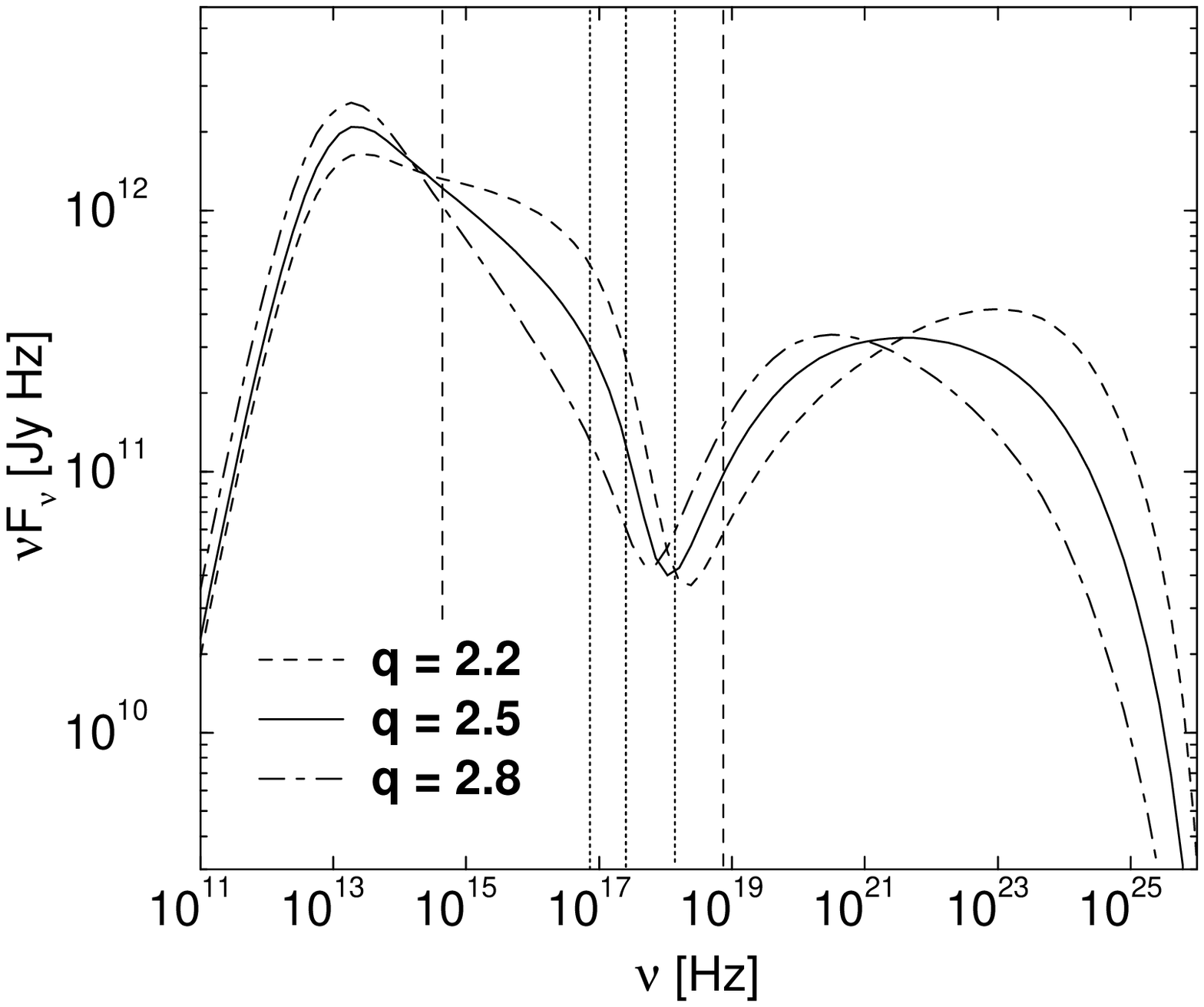}
\caption{Time averaged photon spectra for different values of the 
electron injection spectral index $q$, from simulations no. 2, 9, and 10.
The dotted vertical lines indicate the frequencies at which light curves
and hardness-intensity correlations have been extracted; the dashed
vertical lines indicate the remaining two frequencies at which light
curves have been extracted.}
\label{q_intspectra}
\end{figure}

\begin{figure}
\plotone{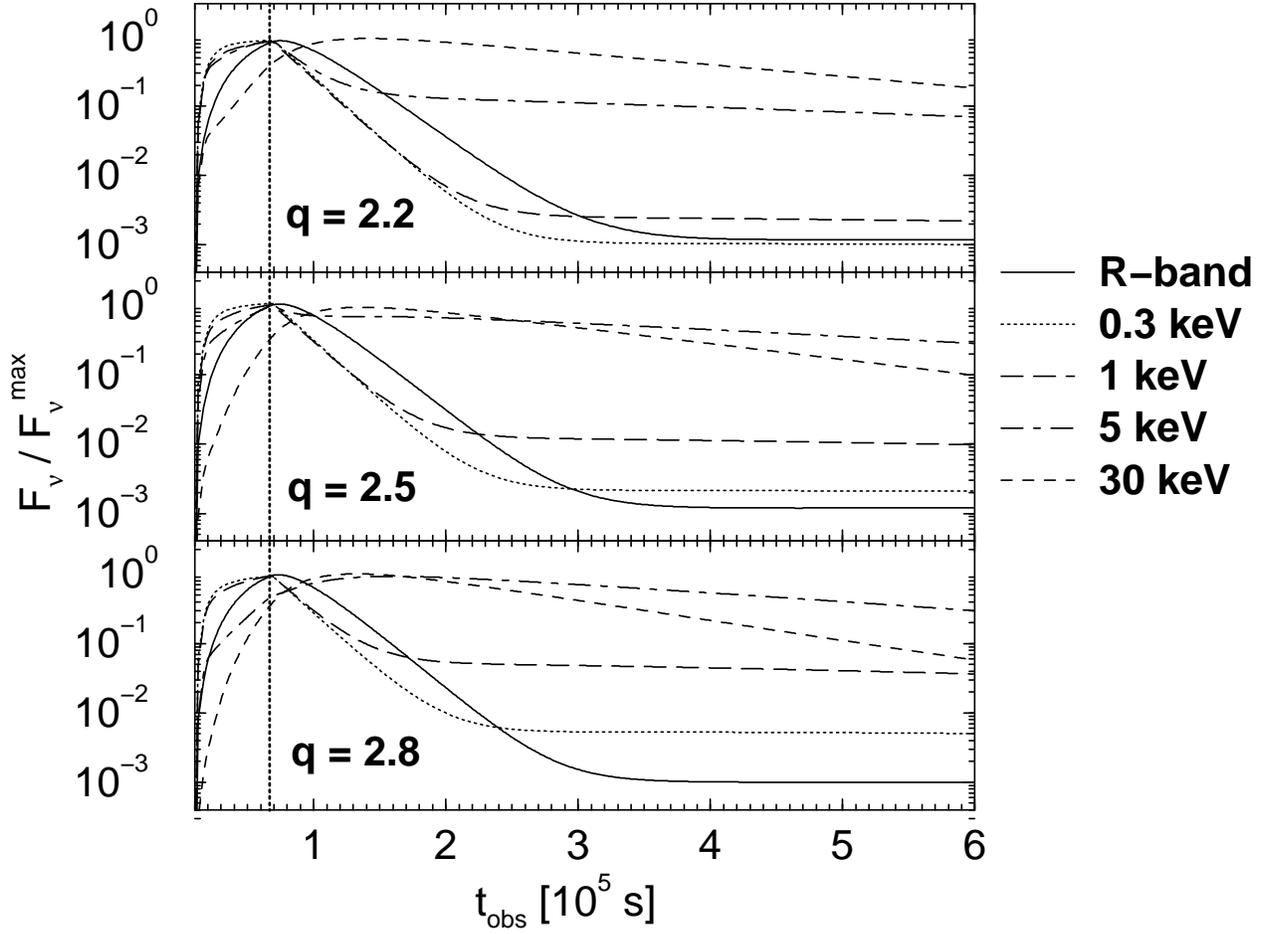}
\caption{Optical and X-ray light curves for different values of
the electron injection spectral indes $q$, from simulations no. 2, 
9, and 10.}
\label{q_lc}
\end{figure}

\begin{figure}
\plotone{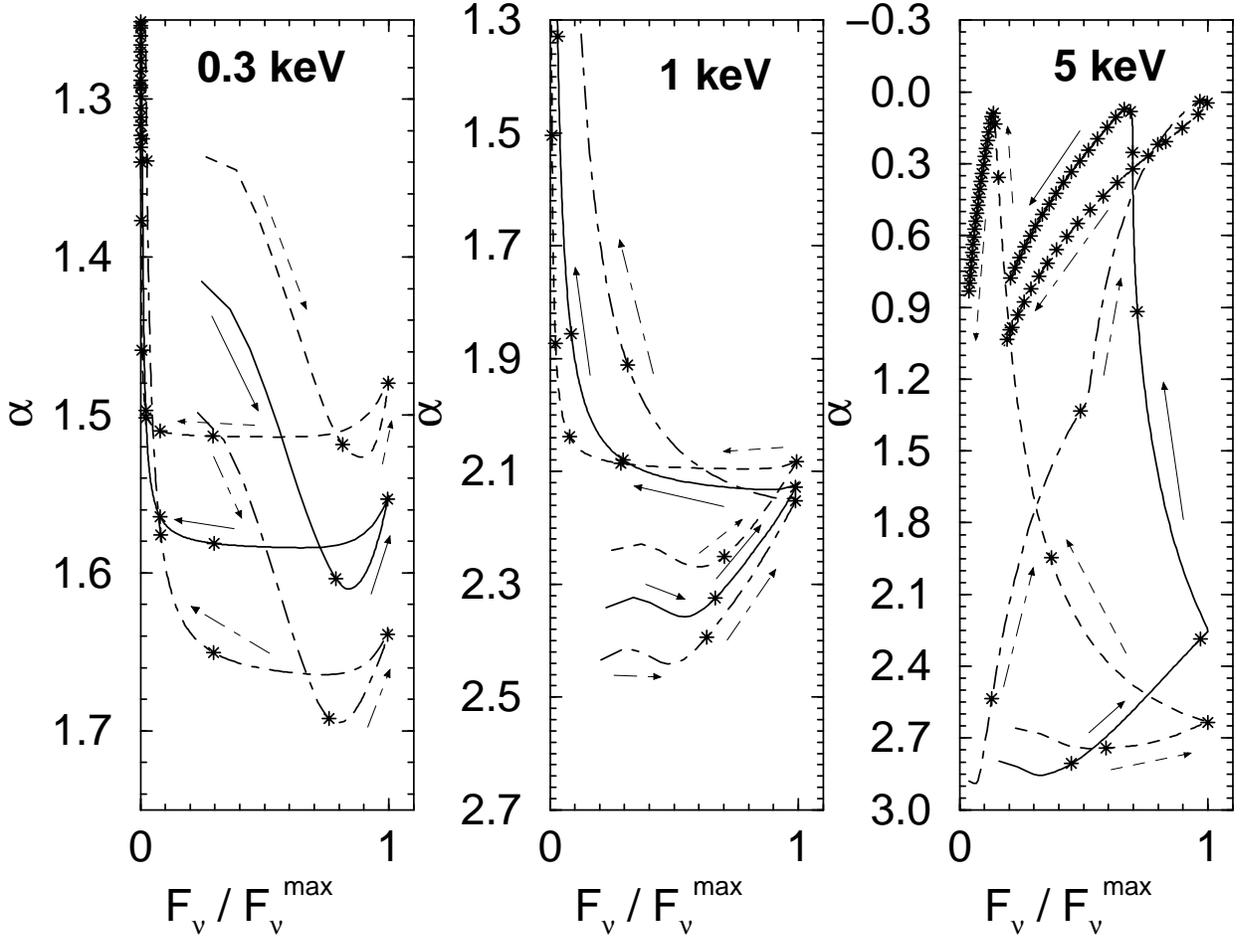}
\caption{Tracks in the harness-intensity diagrams at X-ray energies 
for different values of the electron injection spectral index $q$, from 
simulations no. 2, 9, and 10: dashed ($q = 2.2$), solid ($q = 2.5$),
dot-dashed ($q = 2.8$). Stars indicate the locations at multiples of 
the dynamical time scale during the respective simulations.}
\label{q_hic}
\end{figure}

\begin{figure}
\plotone{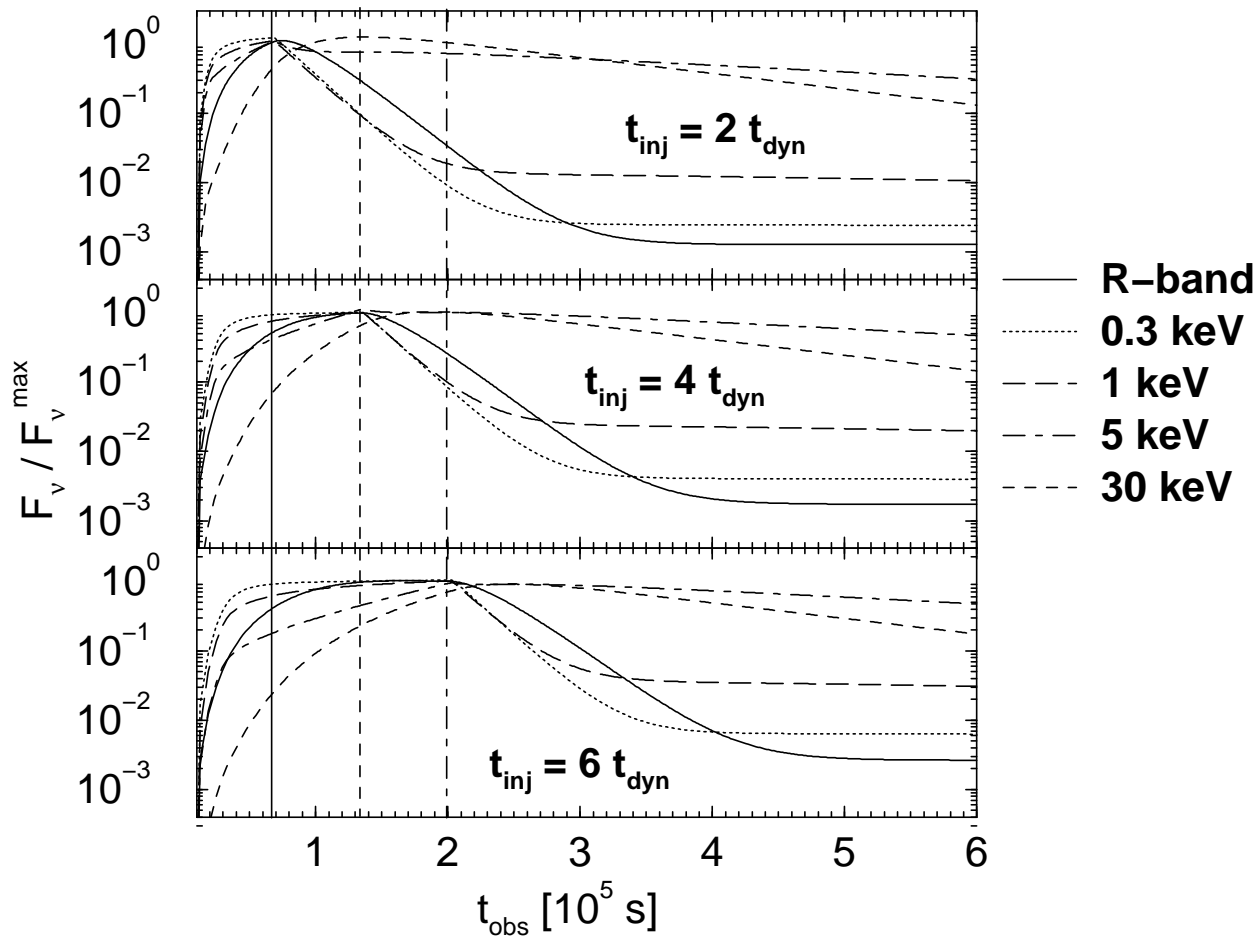}
\caption{Optical and X-ray light curves for different values of
the duration of the flaring injection time scale, keeping the total
energy input during the injection event constant (simulations no. 2, 
11, and 12).}
\label{duration_lc}
\end{figure}

\begin{figure}
\plotone{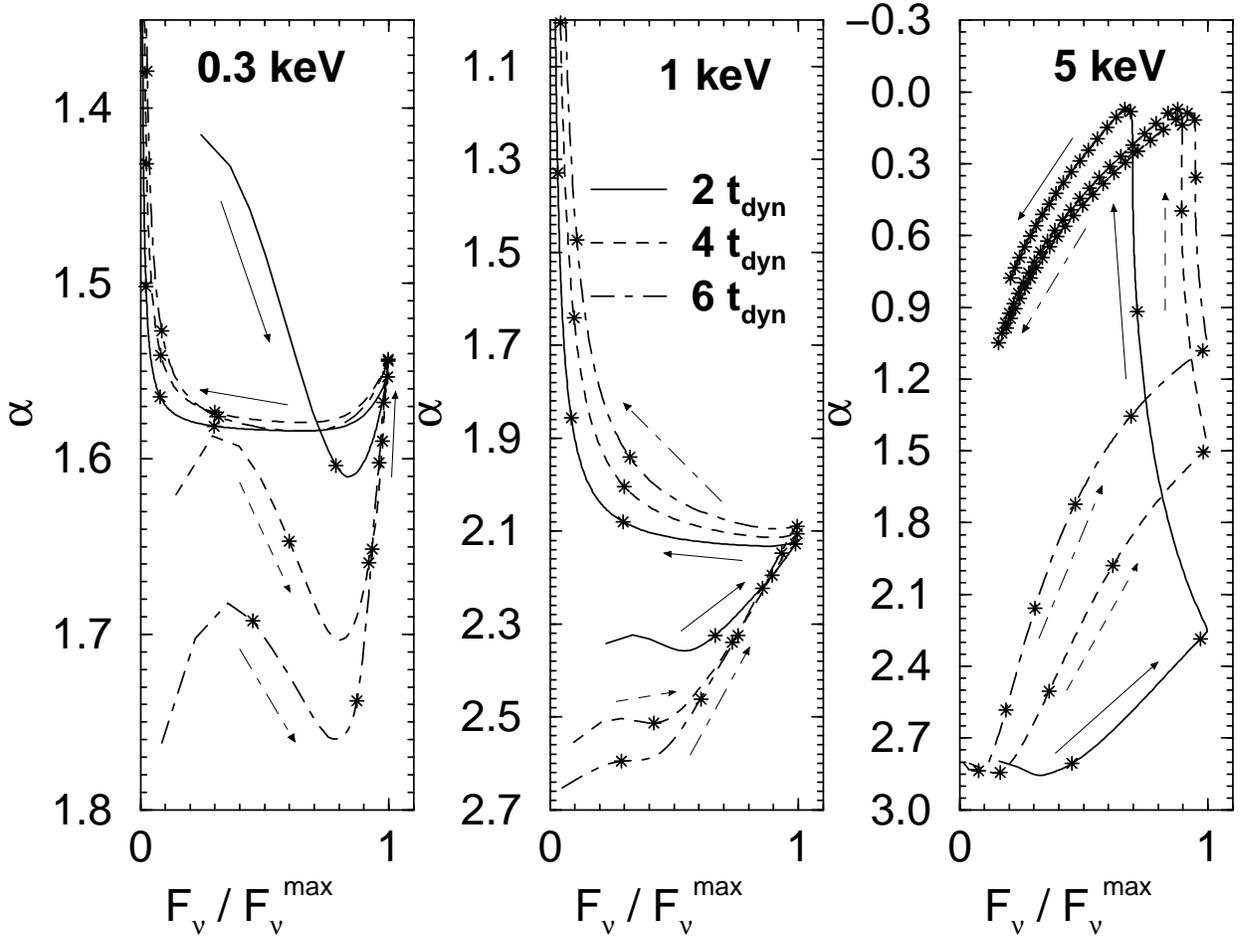}
\caption{Tracks in the harness-intensity diagrams at X-ray energies 
for different values of the flaring injection time scale, keeping the 
total energy input during the injection event constant (simulations 
no. 2, 11, and 12): solid ($t_{\rm inj} = 2 \, t_{\rm dyn}$), dashed 
($t_{\rm inj} = 4 \, t_{\rm dyn}$), dot-dashed ($t_{\rm inj} = 6 \, 
t_{\rm dyn}$). Stars indicate the locations at multiples of 
the dynamical time scale during the respective simulations.}
\label{duration_hic}
\end{figure}

\begin{figure}
\plotone{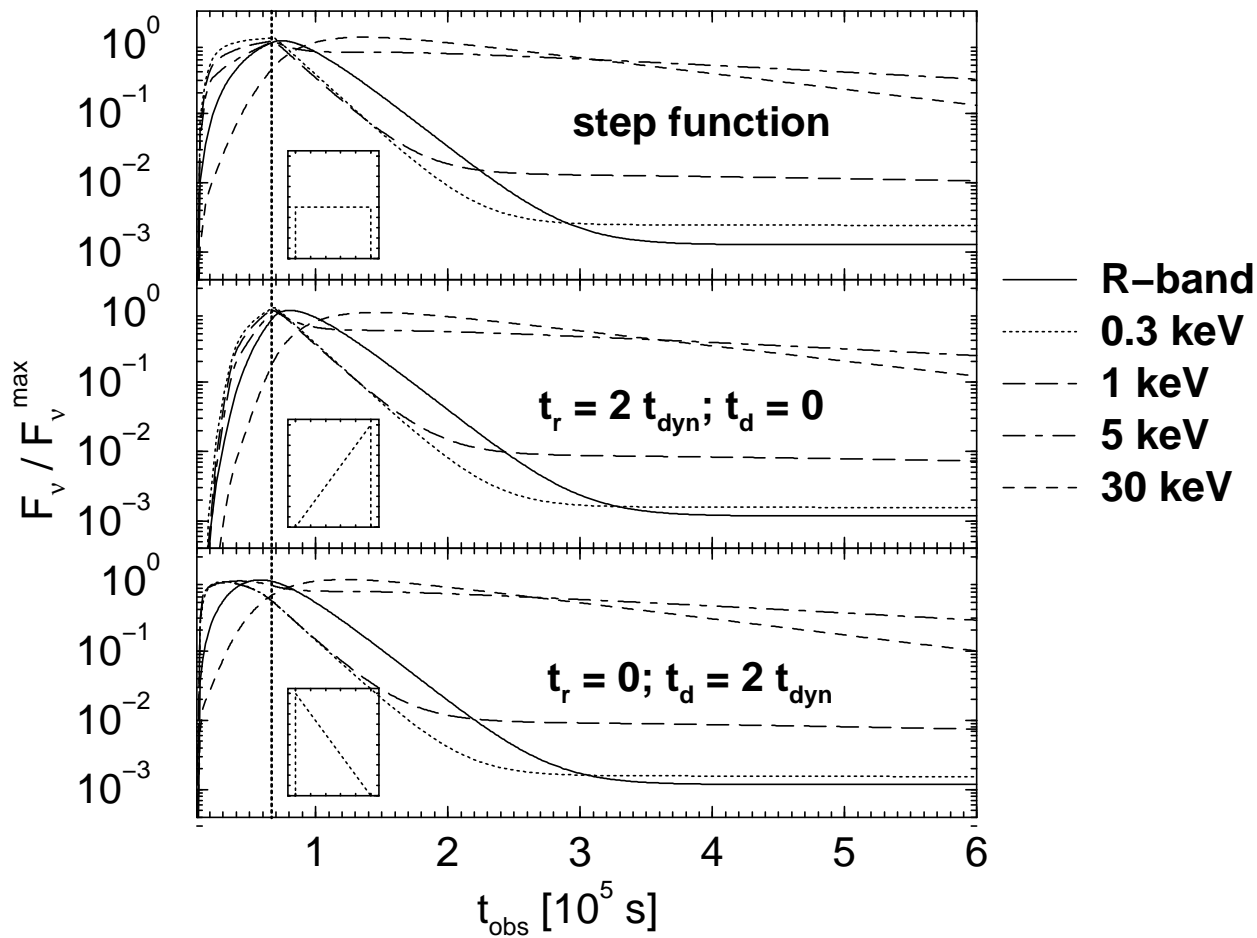}
\caption{Optical and X-ray light curves for different time profiles
of the electron injection power, keeping the total energy input during 
the injection event constant (simulations no. 2, 14, and 15). The small
insets illustrate the injection time profiles ($L_{\rm inj}$ vs. time).}
\label{profile_lc}
\end{figure}

\begin{figure}
\plotone{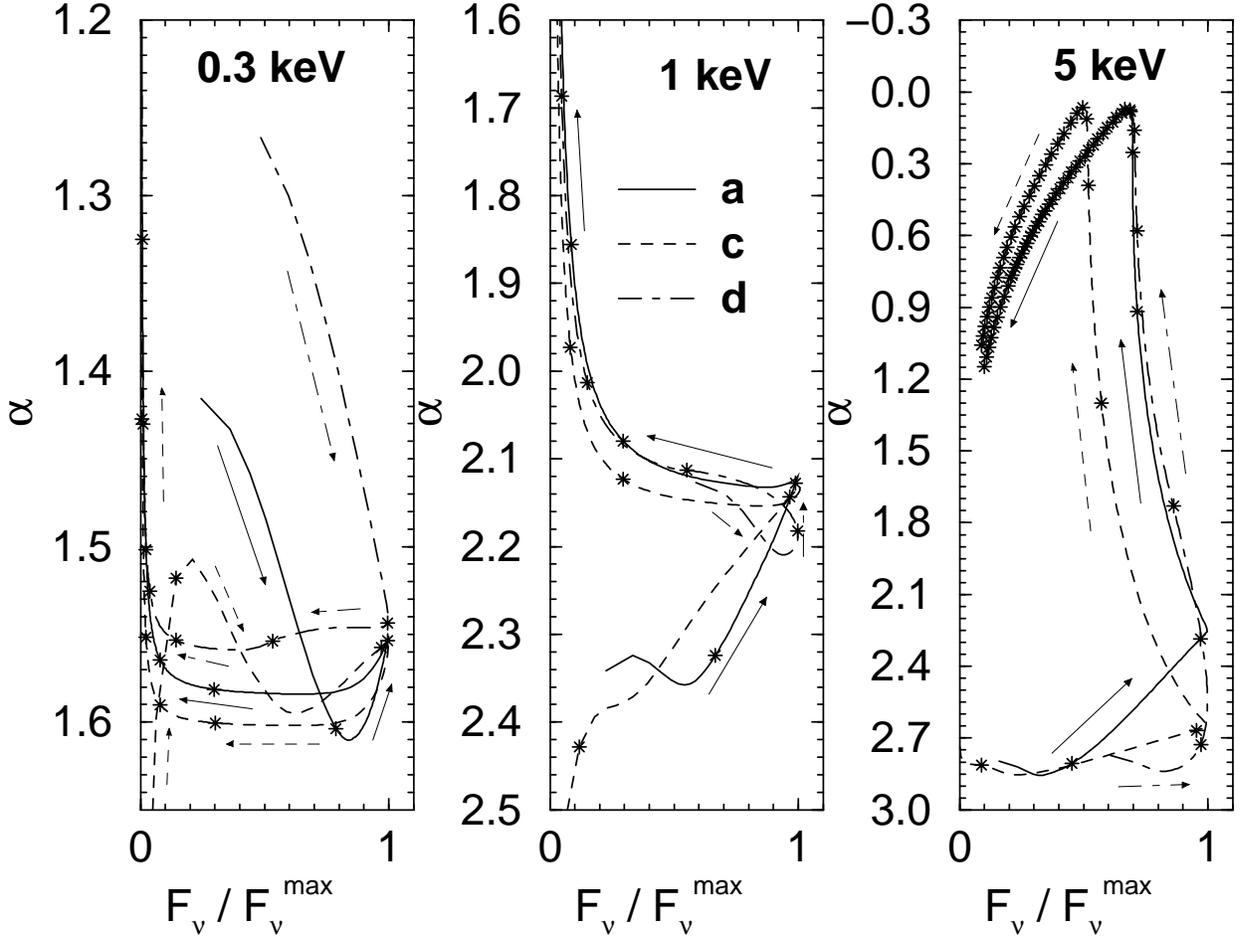}
\caption{Tracks in the harness-intensity diagrams at X-ray energies 
for different time profiles of the electron injection power, keeping 
the total energy input during the injection event constant (simulations 
no. 2, 14, and 15): solid (profile a = step function), dashed (profile c
= triangular profile with $t_r = 2 \, t_{\rm dyn}$ and $t_d = 0$), 
dot-dashed (profile d = triangular profile with $t_r = 0$ and $t_d 
= 2 \, t_{\rm dyn}$). Stars indicate the locations at multiples of 
the dynamical time scale during the respective simulations.}
\label{profile_hic}
\end{figure}

Very important conclusions can be drawn from a comparison of our
results concerning $\gamma$-ray bright sources dominated by either
SSC or external Compton emission. For a given level of flux at
GeV energies, at a level comprable to or exceeding the synchrotron
$\nu F_{\nu}$ peak flux, those two scenarios should be clearly 
distinguishable by virtue of the X-ray variability during flaring 
episodes: If roughly symmetric flare time profiles at soft X-rays
below the synchrotron cutoff are observed, and the local, time-resolved 
X-ray spectra are soft --- consistent with a case with negligible
$\gamma$-ray emission --- the $\gamma$-ray emission might be dominated
by external-Compton emission (see Figs. \ref{ext_intspectra} 
-- \ref{ext_hic}). However, if the X-ray time profiles
show a clear sign of a very rapid rise and more gradual decay, and
the time-resolved spectra are significantly harder than corresponding
to a case without strong $\gamma$-ray emission, we expect a strong
contribution from the SSC mechanism to the $\gamma$-ray emission
(see Figs. \ref{L_inj_intspectra} -- \ref{L_inj_hic}). Most 
notably, this diagnostic of the $\gamma$-ray emission mechanism 
does not require a detailed spectral measurement --- which will be
hard to achieve and require long integration times, even with GLAST 
---, but only a rough estimate of the GeV flux. It relies primarily 
on X-ray variability studies, combined with the type of detailed 
time-dependent radiation modeling presented in this paper. 

We have shown that our results are not significantly impacted by
the special choice of poorly determined parameters like the details
(exact duration and time profile) of the acceleration / injection
events leading to flaring activity, or the electron escape time scale
parameter. Consequently, the X-ray variability of the high-frequency
end of the synchrotron emission in intermediate and low-frequency
peaked BL~Lac objects can be used as a very robust diagnostic to
unveil the nature of the high-energy emission in this type of blazars. 

\acknowledgements 
We thank C. D. Dermer for discussions and very detailed and useful 
comments, and the anonymous referee for a careful review and a helpful 
report. The work of MB was supported by NASA through {\it Chandra} 
Postdoctoral Fellowship Award no. 9-10007, issued by the 
{\it Chandra} X-ray Center, which is operated by the Smithsonian 
Astrophysical Observatory for and on behalf of NASA under 
contract NAS~8-39073.

\newpage

\newpage

\newpage

\newpage

\newpage

\newpage

\newpage

\newpage

\newpage

\newpage

\newpage

\newpage

\newpage

\newpage

\newpage

\newpage

\newpage

\newpage

\begin{deluxetable}{ccccccccc}
\tabletypesize{\scriptsize}
\tablecaption{Parameters of the simulations used for our parameter study. Throughout the series of
simulations, we have used $L_{\rm inj}^{\rm qu} = 10^{38}$~ergs~s$^{-1}$ (electron injection luminosity
during quiescence), $\gamma_{1} = 10^3$ (low-energy cutoff of injected electron spectrum), $\gamma_2 
= 10^5$ (high-energy cutoff of injected electron spectrum), $D = 10$ (Doppler boosting factor), 
$R_b = 10^{16}$~cm (blob radius), and $\epsilon_B = 1$ (magnetic-field equipartition parameter). 
The time profiles in the 6th column are: (a) step function, (b) triangular with linear rise and
decay with equal time scales ($t_{\rm r,d} = t_{\rm inj}/2$), (c) linear rise and instantaneous 
drop ($t_{\rm r} = t_{\rm inj}$), (d) instantaneous rise and linear decay ($t_{\rm d} = t_{\rm inj}$).
For time profiles (b) -- (d), the parameter $L_{\rm inj}^{\rm fl}$ is the maximum injection power.
$\eta$ is the electron escape time parameter, defined by $t_{\rm e, esc} = \eta \, R_b / c$. The
parameters changed with respect to the base model (no. 2) printed in boldface.}
\tablewidth{0pt}
\tablehead{
\colhead{Run no.} & \colhead{$L_{\rm inj}^{\rm fl}$ [ergs s$^{-1}$]} & \colhead{$\tau_{\rm T, BLR}$} & 
\colhead{$q$} & \colhead{$\Delta t_{\rm inj}$} & \colhead{inj. time profile} & \colhead{$\eta$}
}
\startdata
1 & {\bf 10}$^{\bf 40}$ & 0 & 2.5 & $2 \, t_{\rm dyn}$ & a & 10 \\
2 & $10^{41}$ & 0 & 2.5 & $2 \, t_{\rm dyn}$ & a & 10 \\
3 & {\bf 10}$^{\bf 42}$ & 0 & 2.5 & $2 \, t_{\rm dyn}$ & a & 10 \\
4 & $10^{41}$ & {\bf 10}$^{\bf -3}$ & 2.5 & $2 \, t_{\rm dyn}$ & a & 10 \\
5 & $10^{41}$ & {\bf 10}$^{\bf -2}$ & 2.5 & $2 \, t_{\rm dyn}$ & a & 10 \\
6 & $10^{41}$ & {\bf 10}$^{\bf -1}$ & 2.5 & $2 \, t_{\rm dyn}$ & a & 10 \\
7 & $10^{41}$ & {\bf 1} & 2.5 & $2 \, t_{\rm dyn}$ & a & 10 \\
8 & {\bf 10}$^{\bf 43}$ & 0 & 2.5 & $2 \, t_{\rm dyn}$ & a & 10 \\
9 & $10^{41}$ & 0 & {\bf 2.2} & $2 \, t_{\rm dyn}$ & a & 10 \\
10 & $10^{41}$ & 0 & {\bf 2.8} & $2 \, t_{\rm dyn}$ & a & 10 \\
11 & ${\bf 5 \times 10}^{\bf 40}$ & 0 & 2.5 & {\bf 4 t}$_{\bf dyn}$ & a & 10 \\
12 & ${\bf 3.33 \times 10}^{\bf 40}$ & 0 & 2.5 & {\bf 6 t}$_{\bf dyn}$ & a & 10 \\
13 & $2 \times 10^{41}$ & 0 & 2.5 & $2 \, t_{\rm dyn}$ & {\bf b} & 10 \\
14 & $2 \times 10^{41}$ & 0 & 2.5 & $2 \, t_{\rm dyn}$ & {\bf c} & 10 \\
15 & $2 \times 10^{41}$ & 0 & 2.5 & $2 \, t_{\rm dyn}$ & {\bf d} & 10 \\
16 & $10^{41}$ & 0 & 2.5 & $2 \, t_{\rm dyn}$ & a & {\bf 3} \\
17 & $10^{41}$ & 0 & 2.5 & $2 \, t_{\rm dyn}$ & a & {\bf 30} \\
18$^*$ & $10^{41}$ & {\bf 1} & 2.5 & $2 \, t_{\rm dyn}$ & a & 10 \\
\enddata
\tablenotetext{*}{In this simulation, the magnetic field was held constant 
at $B = 0.4$~G.}\label{parameters}
\end{deluxetable}

\end{document}